\documentclass[runningheads]{llncs}

\usepackage[paperheight=235mm, paperwidth=155mm,textwidth=12.2cm,textheight=19.3cm]{geometry}
\usepackage{amsmath}
\usepackage{styles/ProCount}
\usepackage{styles/tensors}
\usepackage{styles/smallsec}

\usepackage{amsthm}
\usepackage{graphicx}
\usepackage{tikz}
\usepackage{mathtools}
\usepackage{amssymb}
\usepackage{xcolor}
\usepackage[T1]{fontenc}
\usepackage{subcaption}
\usepackage{stmaryrd}
\usepackage[]{algorithmicx}
\usepackage{algpseudocode}
\usepackage{todonotes}
\usepackage{bbding}
\usepackage{float}
\title{Dynamic Programming for Symbolic Boolean Realizability and Synthesis\protect\footnote{This is the full version of this paper, with appendices included.}}

\author{Yi Lin \inst{1}\orcidID{0000-0001-8443-2246} \and 
Lucas M. Tabajara \inst{1}\orcidID{0000-0001-9608-1404}\thanks{Currently at Runtime Verification, Inc.} \and
Moshe Y. Vardi \Envelope \inst{1}\orcidID{0000-0002-0661-5773}}
\authorrunning{Y. Lin et al.}
\institute{Rice University, Houston TX 77005, USA \\
\email{vardi@cs.rice.edu, l.martinelli.tabajara@gmail.com, yl182@rice.edu}}
\begin{document}

\maketitle



\begin{abstract}
Inspired by recent progress in dynamic programming approaches for weighted model counting, we investigate a dynamic-programming approach in the context of boolean realizability and synthesis, which takes a conjunctive-normal-form boolean formula over input and output variables, and aims at synthesizing witness functions for the output variables in terms of the inputs. We show how graded project-join trees, obtained via tree decomposition, can be used to compute a BDD representing the realizability set for the input formulas in a bottom-up order. We then show how the intermediate BDDs generated during realizability checking phase can be applied to synthesizing the witness functions in a top-down manner.
    An experimental evaluation of a solver -- DPSynth -- based on these ideas demonstrates that our approach for Boolean realizabilty and synthesis has superior time and space performance over a heuristics-based approach using same symbolic representations. We discuss the advantage on scalability of the new approach, and also investigate our findings on the performance of the DP framework.
\keywords{Boolean synthesis \and Binary decision diagram \and Dynamic programming 
\and Bucket elimination.}

\end{abstract}
\section{Introduction}\label{sec: introduction(new)}
\noindent











\noindent The \emph{Boolean-Synthesis Problem} \cite{hachtel2007logic} -- a fundamental problem in computer-aided design -- is the problem of taking in a declarative boolean relation between two sets of boolean variables -- input and output --  and generating boolean functions, called \emph{witness functions}, that yield values to the output variables with respect to the input variables so as to satisfy the boolean relation. As a fundamental problem in computer-aided design, there are many applications of boolean synthesis in circuit design. For example, based on a circuit’s desired behavior, we can use boolean synthesis to automatically construct the missing components of the circuit~\cite{akshay2021boolean}. In addition to being used in circuit design, boolean synthesis has recently found applications also in temporal synthesis \cite{ZhuTLPV17,ZhuTLPV17a}, where the goal is to construct a sequential circuit that responds to environment inputs in a way that is guaranteed to satisfy given temporal-logic specification.

Many approaches have been investigated for boolean synthesis, such as knowledge compilation~\cite{akshay2019knowledge}, QBF solving~\cite{DBLP:conf/sat/RabeS16}, and machine learning~\cite{golia2020manthan,manthan2}.
Here we build on previous work on boolean synthesis using Binary Decision Diagrams (BDDs) \cite{FMCAD17}. 
The main advantage of the decision-diagram approach is that it provides not just the synthesized witness functions, but also the realizability set of the specification, which is the set of inputs with respect to which the specification is realizable. In \emph{modular circuit design} \cite{hachtel2007logic}, where a system is composed of multiple modules that are independently constructed, it is imperative to confirm the realizability set of a module, as it has to match the output set of prior modules. Similarly, in the context of temporal synthesis \cite{ZhuTLPV17,ZhuTLPV17a}, the winning set is constructed iteratively by taking the union of realizability sets, where for each realizability set we construct a witness function. The winning strategy is then constructed by stitching together all these witness functions.  

Motivated by applications that require the computation of both realizability sets and witness functions, we describe here a decision-diagram approach that can also handle \emph{partially realizable} specifications, where the realizability set is neither empty nor necessarily universal. (Indeed, in our benchmark suit, about 30\% of the benchmarks are partially realizable.) Our tool, \emph{DPSynth} computes the realizability sets and witness functions with respects to these sets.
While several recent synthesis  tools do provide witnesses for partially-realizable specifications, cf. \cite{akshay2019knowledge,golia2020manthan,manthan2}, not all of them directly output the realizability set, requiring it, instead, to be computed from the witnesses and the original formula.

The boolean-synthesis problem starts with a boolean  formula $\varphi(X,Y)$ over sets $X,Y$ of \emph{input} and \emph{output} variables. The goal is to construct boolean formulas, called \emph{witness functions} (sometimes called  \emph{Skolem functions})~\cite{manthan2,akshay2019knowledge} -- for the \emph{output variables}  expressed in terms of the \emph{input variables}. The BDD-based approach to this problem \cite{CAV16} constructs the BDD $B_\varphi(X,Y)$ for $\varphi$, and then quantifies existentially over the $Y$ variables to obtain a BDD over the $X$ variables that captures the \emph{realizability set} -- the set of assignments $\tau_X$ in $2^X$ for which an assignment $\tau_Y$ in $2^Y$ exists where $B_\varphi(\tau_X,\tau_y)=1$. The witness functions can then be constructed by iterating over the intermediate steps of the realizability computation \cite{CAV16}.

A challenge of this BDD-based approach is that it is often infeasible to construct the BDD $B_\varphi$. \emph{Factored Boolean Synthesis} \cite{FMCAD17} assumes that the formula $\phi$ is given in conjunctive normal form (CNF), where the individual clauses are called \emph{factors}. Rather than constructing the monolithic BDD $B_\varphi$, this approach constructs a BDD for each factor, and then applies conjunction in a lazy way and existential quantification in an eager way, using various heuristics to order conjoining and quantifying. As shown in \cite{FMCAD17}, the factored approach, \emph{Factored RSynth}, is more scalable than the monolithic approach, RSynth. Thus, we use here Factored RSynth as the baseline for comparison in this paper.

By \emph{dynamic programming} we refer to the approach that simplifies a complicated problem by breaking it down into simpler sub-problems in a recursive manner \cite{bellman1966dynamic}. Unlike search-based approaches to Boolean reasoning, cf.~\cite{Moskewicz2001chaff}, which directly manipulate truth assignments, we use data structure based on decision diagrams \cite{bryant1986graph} that represents sets of truth assignments. This approach is referred to as \emph{symbolic}, going back to \cite{burch1992symbolic}. 

The \emph{symbolic dynamic-programming} approach we propose here is inspired by progress in \emph{weighted model counting}, which is the problem of counting the number of satisfying (weighted) assignments of boolean formulas. Dudek, Phan, and Vardi proposed in \cite{dudek2020addmc}, an approach based on {\em Algebraic Decision Diagrams}, which are the quantitative variants of BDDs \cite{bahar1997algebric}. Dudek et al. pointed out that a monolithic approach is not likely to be scalable, and proposed a factored approach, ADDMC, analogous to the approach in \cite{FMCAD17}, in which conjunction is done lazily and quantification eagerly. In follow-up work \cite{dudek2020dpmc}, they proposed a more systematic way to order the quantification and conjunction operations, based on dynamic programming over \emph{project-join trees}; the resulting tool, DPMC, was shown to scale better than ADDMC. In further follow-on work, they proposed \emph{graded project-join trees} for \emph{projected model counting}, where the input formula has two sets of variables -- quantified variables and counting variables \cite{Procount}. Graded project-join tree offers a recursive decomposition of the Boolean-Synthesis Problem into smaller tasks of projections and join. 



We show here how symbolic dynamic programming over graded project-join trees can be applied to boolean synthesis. In our approach, realizability checking is done using a BDD-based \emph{bottom-up} execution, analogous to the handling of counting quantifiers in \cite{Procount}. This enables us to compute the realizability set and check its nonemptiness. We then present a novel algorithm for synthesizing the witness functions using \emph{top-down} execution on graded project-join trees. (In contrast, \cite{Procount} both types of quantifiers are handled bottom-up). We demonstrate the advantage of our bottom-up-top-down approach by developing a tool, \emph{DPSynth}. For a fair evaluation, we compare its performance to Factored RSynth, which can also handle partially realizable specifications. 

The main contributions of this work are as follows. First, we show how to adapt the framework of projected counting to Boolean synthesis. In projected counting there are two types of existential quantifiers -- additive and disjunctive \cite{Procount}, while Boolean synthesis combine universal and existential quantifiers. Second, projected counting requires only a bottom-up pass over the graded project-join trees, while here we introduce an additional, top-down pass over the tree to perform the major part of boolean synthesis, which is,witness construction.

The organization of the paper is as follows.
After preliminaries in Section \ref{sec: preliminaries(new)}, we show in Section \ref{sec: realizability(new)} how to use graded project-join trees for boolean realizability checking, and show that DPSynth generally scales better than Factored RSynth.
Then, in Section \ref{sec: synthesis(new)}, we show how to extend this approach from realizability checking to witness-function construction. Experimental evaluation shows that DPSynth generally scales better than Factored RSynth also for witness-function construction. We offer concluding remarks in Section~\ref{sec:conclusions}.

\section{Preliminary Definitions}\label{sec: preliminaries(new)}

\subsection{Boolean Formula and Synthesis Concepts} \noindent A \emph{boolean formula} $\phi(X)$, over a set $X$ of variables, represents a \emph{boolean function} $f: 2^{X} \to \B$, which selects subsets of $2^{X}$. A truth assignment $\tau$ \emph{satisfies} $\phi$ iff $\phi(\tau) = 1$. A boolean formula in \emph{conjunctive normal form (CNF)} is a conjunction of \emph{clauses}, where a clause is a disjunction of \emph{literals} (a boolean variable or its negation). When $\phi$ is in CNF, we abuse notation and also use $\phi$ to denote its own set of clauses.
Given a CNF formula $\phi(X,Y)$ over \emph{input} and {\em output} variable $X$ and $Y$, the \emph{realizability set} of $\phi$, denoted $R_\phi(X) \subseteq 2^X$, is the set of assignments $\sigma \in 2^X$ for which there exists an assignment $\tau \in 2^Y$ such that $\varphi(\sigma \cup \tau) = 1$. When $\phi$ is clear from context, we simply denote the realizability set by $R$.

\begin{definition}[Realizability]\label{def: full real}\label{def: partial real}\label{def: null real}
    Let $\phi(X,Y)$ be a CNF formula with $X$ and $Y$ as input and output variables. We say that $\phi$ is \emph{fully realizable} if $R = 2^X$. We say that $\phi$ is \emph{partially realizable} if $R \neq \emptyset$. Finally, we say that $\phi$ is \emph{nullary realizable} if $R = \emptyset$.
\end{definition}

\noindent Given the condition that the formula is at least partially realizable, 
the boolean synthesis problem asks for a set of \emph{witnesses} for the output variables generated on top of given input values, such that the formula is satisfied. This sub-problem of constructing witnesses, is usually referred to as synthesis. 

The motivation behind synthesizing partially-realizable specifications is that there are cases where a specification is not fully realizable, but it is still useful to synthesize a function that works for all inputs in the realizability set. An example is the factorization benchmark family, as discussed in the introduction of~\cite{akshay2021boolean}, which takes an integer and aims to factor it into two integers that are both not equal to 1. If the integer is prime, then there is no valid factorization for this particular input, but it would still be valuable to have a solution that works for all composite numbers. This is why our attention to partially-realizable cases is a contribution of the paper, while related works tend to focus mostly on fully-realizable instances.

\begin{definition}[Witnesses in Boolean Synthesis Problem]\label{def: witness construction}
    Let $\phi(X,Y)$ denote a fully or partially realizable boolean formula  with input variables in $X=\{x_1,\ldots,x_m\}$ and output variables $Y =\{y_1,\ldots,y_n\}$, and let $R_{\phi}(X) \neq \emptyset$ be its realizability set. A sequence  $g_1(X), \ldots, g_n(X)$ of boolean functions is a sequence of \emph{witness functions} for the $Y$ variables in $\phi(X,Y)$ if for every assignment ${x} \in R_{\phi}(X)$, we have that
    $\phi[X \mapsto {x}][y_1 \mapsto g_1({x})]\ldots[y_n \mapsto g_n({x})]$ holds.
\end{definition}

\begin{definition}[Synthesis]\label{def: synthesis}
     Given a partially or fully realizable CNF formula $\phi(X,Y)$ with input and output variables $X$ and $Y$, the \emph{synthesis problem} asks to algorithmically construct a set of \emph{witness functions} for the $Y$ variables in terms of the $X$ variables.     
\end{definition}

\subsection{Dynamic Programming Concepts - Project-Join Trees}

A \emph{binary decision diagrams (BDD)}~\cite{bryant1986graph} is directed acyclic graphs with two terminals labeled by $0$ and $1$. A BDD provides a canonical representation of boolean functions (and, by extension, boolean formulas). A (reduced, ordered) BDD is constructed from a binary decision tree, using a uniform variable order, of a boolean function by merging identical sub-trees and suppressing redundant nodes (nodes where both children are the same). Each path from the root of the BDD to the 1-terminal represents a satisfying assignment of the boolean function it represents. Based on the combination of BDDs and project-join trees, which is to be defined below, our dynamic-programming algorithm solves the boolean synthesis problem.


\begin{definition}[Project-Join Tree of a CNF formula]\label{def: project-join tree}
A \emph{project-join tree}~\cite{dudek2020dpmc} for a CNF formula $\phi(X)$ is defined as a tuple $\T = (T, r, \gamma, \pi)$, where 
\begin{enumerate}
    \item $T$ is a tree with a set $\V{T}$ of vertices, a set $\Lv{T} \subseteq \V{T}$ of leaves, and a root $r \in \V{T}$
    \item $\gamma : \Lv{T} \to \phi$ is a bijection that maps the leaves of $T$ to the clauses of $\phi$
    \item $\pi : \V{T} \setminus \Lv{T} \to 2^X$ is a function which labels internal nodes with variable sets, where the labels $\{\pi(n) \mid n \in \V{T} \setminus \Lv{T}\}$ form a partition of $X$,
    and \item If a clause $c \in \phi$ contains a variable $x$ that belongs to the label $\pi(n)$ of an internal node $n \in \V{T} \setminus \Lv{T}$, then the associated leaf node $\gamma^{-1}(c)$ must descends from $n$.
\end{enumerate} 
\end{definition}



A \emph{Graded Project-Join Tree}~\cite{Procount} is a generalization of project-join trees.
\begin{definition}[Graded Project-Join Tree of a CNF formula~\cite{Procount}]\label{def: graded tree}
    A project-join tree $\T = (T, r, \gamma, \pi)$ of a CNF formula $\phi(X,Y)$ over variables $X\cup Y$, where $X\cap Y=\emptyset$, is \emph{($X$, $Y$)-graded} if there exist \emph{grades} $\I_X, \I_Y \subseteq \V{T}$ that partition the internal nodes $\V{T} \setminus \Lv{T}$, such that:
    \begin{enumerate}
        \item For a node, its grade is always consistent with its labels. i.e., If $n_X \in \I_X$ then $\pi(n_X) \subseteq X$, and if  $n_Y \in \I_Y$ then $\pi(n_Y) \subseteq Y$.
        \item If $n_X \in \I_X$ and $n_Y \in \I_Y$, then $n_X$ is not a descendant of $n_Y$ in $T$.
    \end{enumerate}
\end{definition}

\noindent Intuitively, in a graded project-join tree, the nodes are partitioned according to a partition $(X,Y)$ of the variables, with nodes in the $X$ block always appearing higher than nodes in the $Y$ block. As we shall see, this is useful for formulas with two different types of quantifiers.

\begin{figure}[H]\centering
\begin{subfigure}{0.52\textwidth}\centering
\includegraphics[width=\textwidth]{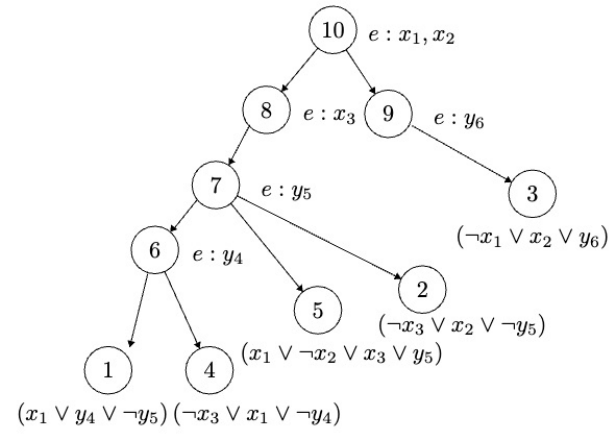}
\caption{A Graded Project-Join Tree Instance} \label{figure: graded tree example}
\end{subfigure}
\hfill
\begin{subfigure}{0.46\textwidth}\centering
\includegraphics[width=\textwidth]{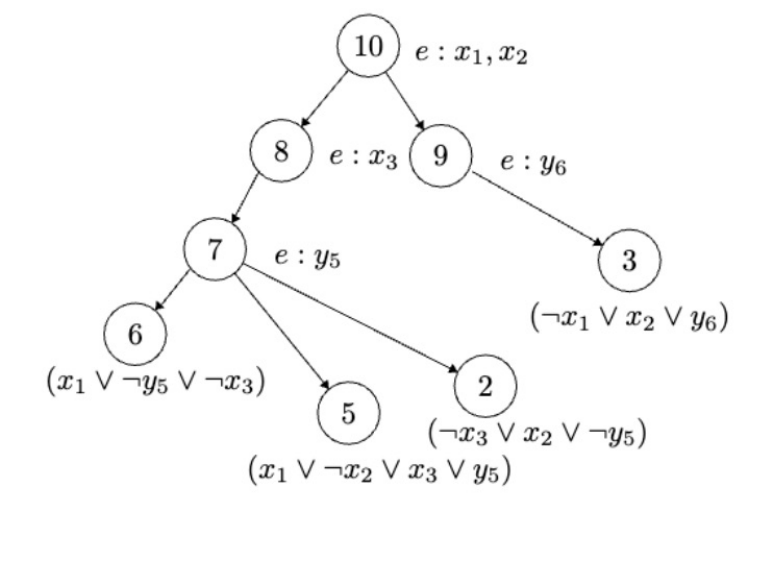}
\caption{Tree after evaluating node 6}
\label{subfigure: intermediate tree 1}
\end{subfigure}

\vfill

\begin{subfigure}{0.5\textwidth}
\includegraphics[width=0.8\textwidth]{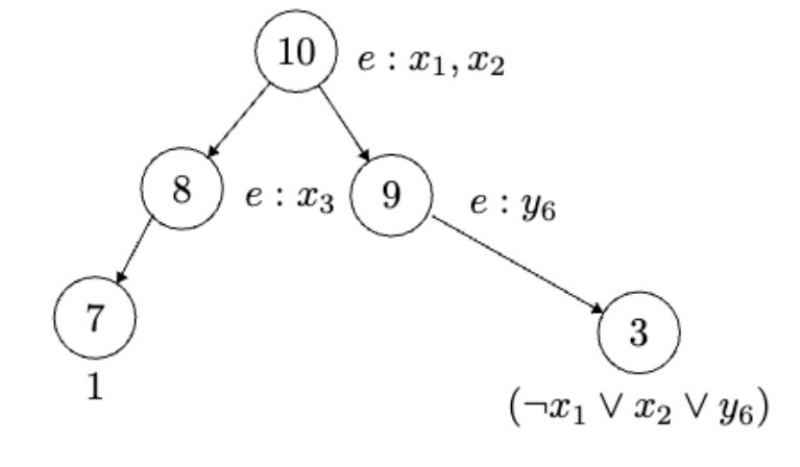}
\caption{Tree after evaluating node 7}
\label{subfigure: intermediate tree 2}
\end{subfigure}
\hfill
\begin{subfigure}{0.45\textwidth}\centering
\includegraphics[width=0.66\textwidth]{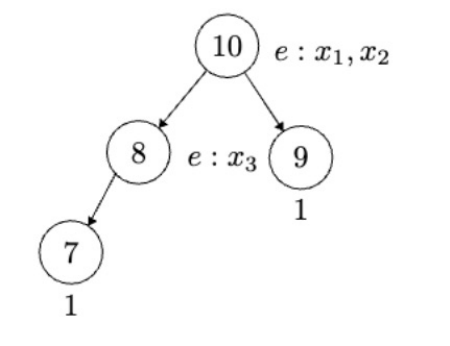}
\caption{Tree after evaluating node 9}
\label{subfigure: intermediate tree 3}
\end{subfigure}
\caption{Original and intermediate trees generated by our algorithms for the CNF example $(x_1 \lor y_4 \lor \neg y_5) \land (\neg x_3 \lor x_2 \lor \neg y_5) \land (\neg x_1 \lor x_2 \lor y_6) \land (\neg x_3 \lor x_1 \lor \neg y_4) \land (x_1 \lor \neg x_2 \lor x_3 \lor y_5)$. The label for each internal node is denoted by $e$.}
\label{figure: intermediate trees}
\end{figure}

Figure~\ref{figure: intermediate trees} shows an example Graded Project-Join Tree for a CNF formula, along with intermediate trees produced in the course of the execution of our synthesis algorithm. In the upcoming sections we will refer back to this example to illustrate the individual steps of the algorithm related to each intermediate tree.



\section{Realizability Checking\protect\footnote{Proofs for all lemmas and theorems can be found in the Appendix A.}}\label{sec: realizability(new)}

\noindent 
Our overall approach has three phases: (1) \emph{planning} -- constructing graded project-join trees, (2) \emph{realizability checking}, and (3) \emph{witness-function synthesis}. The focus in this section is on realizability checking.
We construct the realizability set $R_\varphi(X)$ for an input formula $\varphi(X,Y)$, and then use it to check for full and partial realizability, as described in Definition~\ref{def: full real}.


For the planning phase, we use the planner described in \cite{Procount}, which is based on \emph{tree decomposition} \cite{robertson1991graph}. Computing minimal tree decomposition is known to be an NP-hard problem~\cite{bodlaender2023treewidth}, so planning may incur high computational overhead. Nevertheless, tree-decomposition tools are getting better and better.
The planner uses an \emph{anytime} tree-decomposition tool, cf.~\cite{gogate2012complete}, which outputs tree decompositions of progressively lower width. Deciding when to quit planning and start executing is done heuristically. In contrast, Factored RSynth applies a \emph{fixed} set of fast heuristics, which incurs relatively small overhead. We discuss the planning overhead further below.


\subsection{Theoretical Basis and Valuations in Trees}\label{subsec: quantification on variables} 

The realizability set $R_\varphi(X)$ can be interpreted as a constraint over the $X$ variables stating the condition that there exists an assignment to the $Y$ variables that satisfies $\varphi$. In other words, $R_\varphi(X) \equiv (\exists y_1)\ldots (\exists y_n)\phi$, for $Y = \{y_1, \ldots, y_n\}$. 

Therefore, we can construct $R$ from $\varphi=\bigwedge_j \varphi_j$ by existentially quantifying all $Y$ variables. As observed in~\cite{FMCAD17}, however, a clause that does not contain $y_i$ can be moved outside the existential quantifier $(\exists y_i)$. In other words, $(\exists y_i)\bigwedge_j \varphi_j \equiv \bigwedge_{y_i \not\in AP(\varphi_j)}\varphi_j \land(\exists y_i)\bigwedge_{y_i \in AP(\varphi_j)}\varphi_j$, where $AP$ stands for atomic propositions in a formula. This allows us to perform \emph{early quantification} to compute $R$ more efficiently, since applying quantification on the smaller formula $\bigwedge_{y_i \in AP(\varphi_j)}(\varphi_j)$ is likely less expensive computationally than doing so on the full formula $\varphi = \bigwedge_j(\varphi_j)$.

Inspired by a similar observation in~\cite{Procount}, we use the insight that a graded project-join tree can be employed to guide early quantification. Consider the following definition, which allows us to interpret a node $n$ in a project-join tree as a boolean formula:
\begin{definition}[BDD-Valuations of Nodes in Project Join Tree]\label{def: BDD valuation}
    Let $n$ be a node in a graded project-join tree $\T = (T, r, \gamma, \pi)$ with a partition of internal nodes into two grades $\I_X$ and $\I_Y$, as defined in Section~\ref{sec: preliminaries(new)}. 
    Let the set of children nodes of $n$ be denoted by $C(n)$.
    Let $\llbracket \alpha \rrbracket$ denote the BDD encoding a boolean expression $\alpha$, and $\exists_Z f$ denote the existential projection on $f$ with respect to variables in $Z$. 
    We now define a pair of mutually related valuation concepts for nodes in a project-join tree, interpreting their BDD representations.

    The \emph{post-valuation} of $n$ is defined as
    $$\texttt{BV}_{\texttt{post}}(\T, n)=
        \begin{cases}
		\llbracket \gamma(n) \rrbracket, 
                 \hspace{1cm} &\text{if $n$ is a leaf node}\\
            \exists_{\pi(n)}  \texttt{BV}_{\texttt{pre}}(\T, n), 
                 &\text{if $n$ is an internal node.}
		 \end{cases}$$
    The \emph{pre-valuation} of $n$ is defined as
    $$\texttt{BV}_{\texttt{pre}}(\T, n)=
        \begin{cases}
		\llbracket \gamma(n) \rrbracket, 
                 \hspace{1cm} &\text{if $n$ is a leaf node}\\
            \bigwedge_{n' \in C(n)}  \texttt{BV}_{\texttt{post}}(\T, n'), 
                 &\text{if $n$ is an internal node.}
		 \end{cases}$$
\end{definition}


\noindent Intuitively, we evaluate an internal node $n$ by first taking the conjunction of post-valuations of its children and then existentially quantifying the variables $\pi(n)$ in its label. The former step generates a pre-valuation, while the latter produces a post-valuation.
We can safely perform quantification of the variables in the label of $n$ after conjoining, because every internal node $n$ must satisfy the property that all clauses containing variables in $\pi(n)$ are descendants of $n$, by the definition of a project-join tree.

Furthermore, recall that, by the definition of a $(X, Y)$-graded project-join tree, all nodes in $\I_Y$ occur below all nodes in $\I_X$. This allows us to turn Definition~\ref{def: BDD valuation} into a procedure for efficiently computing realizability of the CNF formula $\varphi$ using a graded project-join tree as a guide:

\begin{enumerate}
    \item First, apply Definition~\ref{def: BDD valuation} to the $\I_Y$ nodes of the tree, producing a project-join tree for $R_\varphi (X) \equiv (\exists y_1) \ldots (\exists y_n) \varphi$.
    \item Inspect this tree for full realizability (see below).
    \item Apply Definition~\ref{def: BDD valuation} again in order to check partial realizability (see below).
\end{enumerate}


For the CNF with the graded project-join tree in Figure~\ref{figure: graded tree example}, the pre- and post-valuations for nodes $1,4,5,2,3$ are equivalent to the BDDs representing their original clauses. $\texttt{BV}_{\texttt{pre}}(\T, 6)=\llbracket (x_1 \lor y_4 \lor \neg y_5) 
\land (\neg x_3 \lor x_1 \lor \neg y_4) \rrbracket, \texttt{BV}_{\texttt{post}}(\T, 6)=(\exists y_4)\texttt{BV}_{\texttt{pre}}(\T, 6) = (x_1 \lor \neg y_5 \lor \neg x_3), 
\texttt{BV}_{\texttt{pre}}(\T, 7)=(x_1 \lor \neg x_2 \lor x_3 \lor y_5) \land (\neg x_3 \lor x_2 \lor \neg y_5)  \land (x_1 \lor \neg y_5 \lor \neg x_3), \texttt{BV}_{\texttt{post}}(\T, 7)=(\exists y_5)\texttt{BV}_{\texttt{pre}}(\T, 7) = \llbracket 1 \rrbracket, \texttt{BV}_{\texttt{pre}}(\T, 9)= (\neg x_1 \lor x_2 \lor y_6)$, and $\texttt{BV}_{\texttt{post}}(\T, 9)=(\exists y_6)\texttt{BV}_{\texttt{pre}}(\T, 9) = \llbracket 1 \rrbracket$.

\paragraph{Notations}We denote the set of children nodes and the set of descendants of $n$ by $C(n)$ and $D(n)$. Let $\llbracket \cdot \rrbracket$ denote the BDD encoding of the enclosed expression, and use $\texttt{projn}(B, Z)$ to denote a series of existential projections on BDD $B$ with respect to variables in set $Z$.


\begin{algorithm}[H]
\caption{$GenericValuation(\T, n)$}\label{alg: generic}
    \DontPrintSemicolon
    \SetKwFunction{this}{GenericValuation}
    \SetKwFunction{exist}{Projn}
    \SetKwInOut{Function}{Operation}
    \SetKwInOut{Parameter}{Notation}
    
    

    \KwIn{: an ($(X,Y)$-graded) project-join tree $\T = (T, r, \gamma, \pi)$ of $\phi$, and a particular node $n \in \V{\T}$}
    \KwOut{early determination of nullary realizability, otherwise outputs $\texttt{BV}_{\texttt{post}}(\T, n)$ and $\texttt{BV}_{\texttt{pre}}(\T, n)$} 

    \uIf {\upshape $n \in \Lv{T}$}{
        $\alpha \gets \llbracket \gamma(n) \rrbracket, \texttt{pre-BV}(\T,n) \gets \alpha$, $\texttt{post-BV}(\T,n) \gets \alpha$\tcp{if $n$ is a leaf}

    } \Else {
        $\texttt{post-BV}(\T,n) \gets \llbracket1\rrbracket$, $\texttt{pre-BV}(\T,n) \gets \llbracket1\rrbracket$\;
        \For{$n' \in C(n)$}{
            \If {$\texttt{post-BV}(\T,n') == \llbracket0\rrbracket$\label{line: viable operation val}}{
                $\texttt{pre-BV}(\T,n) \gets \llbracket0\rrbracket$, $\texttt{post-BV}(\T,n) \gets \llbracket0\rrbracket$\;
                \Return{nullary realizable, no further synthesis needed}\;
            } \Else {$\texttt{pre-BV}(\T,n) \gets \texttt{pre-BV}(\T,n) \land \texttt{post-BV}(\T, n')$
            }

            \If {$\texttt{pre-BV}(\T,n) == \llbracket0\rrbracket$}{
                $\texttt{post-BV}(\T,n) \gets \llbracket0\rrbracket$\;
                \Return{nullary realizable, no further synthesis needed}
            }
        }
        $\texttt{post-BV}(\T,n) \gets \exist(\texttt{pre-BV}(\T,n), \pi(n))$ 
    } 
    \Return{$\texttt{pre-BV}(\T,n), \texttt{post-BV}(\T,n)$}

\end{algorithm}

\noindent Algorithm~\ref{alg: generic} presents a procedure to compute the pre and post-valuations of Definition~\ref{def: BDD valuation}. In this algorithm pre-valuations are intermediate BDDs used to compute the post-valuations, but their values are also used in Section~\ref{sec: synthesis(new)} for witness-function synthesis. 
Post-valuations, meanwhile, are used in Section~\ref{subsec: compute R check full/partial alg 2 and 3} to determine if a given instance has a fully, partially, or unrealizable domain.

Note that children nodes are always visited before parent nodes, per Algorithm~\ref{alg: low valuation}. This guarantees that line~\ref{line: viable operation val} in \texttt{GenericValuation} is always viable.
We now assert the correctness of the Algorithm~\ref{alg: generic} by the following theorem.
\begin{theorem}\label{thm: alg.generic correctness}
If a graded project-join tree $\T$ and a particular node $n$ are given,  then (i) $\texttt{BV}_{\texttt{post}}(\T, n)$ and 
(ii) $\texttt{BV}_{\texttt{pre}}(\T, n)$ 
returned by Algorithm~\ref{alg: generic} are as defined in Definition~\ref{def: BDD valuation}. 
\end{theorem}
\begin{theorem}\label{thm: preval(n)=R(X)}
Given a graded project-join tree $\T$ of a CNF formula $\phi$, let ${{X_\texttt{leaves}}(\T)}$ denote the set of highest level nodes $n \in \T$ such that $\pi(n) \subseteq Y_\phi$. Then the realizability set $R_\varphi(X)$ can be represented by the conjunction $\bigwedge_{n \in {X_\texttt{leaves}}(\T)} \texttt{BV}_{\texttt{pre}}(\T, n)$ of BDDs returned by Algorithm~\ref{alg: generic}.
\end{theorem}




\noindent The pair of post and pre valuations defined in this section offer support for both realizability checking and synthesis, respectively. Section~\ref{subsec: compute R check full/partial alg 2 and 3} applies $\texttt{BV}_\texttt{post}$ for realizability checking, and we show in Section~\ref{sec: synthesis(new)} how $\texttt{BV}_\texttt{pre}$ is used in witnesses construction.

\subsection{Determining Nullary, Partial and Full Realizability}\label{subsec: compute R check full/partial alg 2 and 3}


Using the post-valuations of nodes computed in Algorithm~\ref{alg: generic} as the result of projecting variables on the conjunction of children nodes, we can construct the realizability set and determine if it is full, partial or empty. 
In practice, we represent the realizability set as a conjunction of BDDs, where an input is in the set if it satisfies all BDDs in the conjunction.


\paragraph{Notation}Let $\texttt{XLeaves}(\T)$ denote the set of $Y$ internal nodes whose parents are not in $\I_Y$. This set is easily obtainable by means such as graph-search algorithms.

We start by applying the pair of valuations computed in Algorithm~\ref{alg: generic} to the first layer of $\I_Y$ nodes in the graded tree; that is, all internal nodes in $\I_Y$ whose parent is not in $\I_Y$. By replacing these nodes with leaves labeled by their post-valuation, we obtain a project-join tree $\T_X$ for the formula $(\exists y_1)\ldots(\exists y_n) \varphi$, which, as explained in Section~\ref{subsec: quantification on variables}, corresponds to the realizability set. This procedure is implemented in Algorithm~\ref{alg: low valuation}.


In the case of the example formula in Figure~\ref{figure: intermediate trees}, once we obtain the pre- and post-valuations from Algorithm~\ref{alg: generic}, $B_{pureX}=\llbracket 1 \rrbracket$ and $b=1$ from the post-valuations $\texttt{BV}_{\texttt{post}}(\T, 7)$ (Figure~\ref{subfigure: intermediate tree 2}) and $\texttt{BV}_{\texttt{post}}(\T, 9)$ (Figure~\ref{subfigure: intermediate tree 3}) on $\texttt{XLeaves}$. Hence we get full realizability.

Note that the formula $\varphi$ is fully realizable if and only if the conjunction of the leaves in $\T_X$ is $1$, which holds true if and only if all leaves are $1$. Therefore, at the end of Algorithm~\ref{alg: low valuation} we are also able to answer whether $\varphi$ is fully realizable. Note also that if any of the leaves is $0$, then their conjunction is $0$, meaning that $\varphi$ is nullary realizable. Therefore, in some cases it might be possible to detect nullary realizability in this step. If the formula is not fully realizable and nullary realizability is not detected, then $\T_X$ is passed to the next step to test partial realizability.
\setlength{\floatsep}{5pt plus 2pt minus 2pt}
\setlength{\textfloatsep}{5pt plus 2pt minus 2pt}
\setlength{\intextsep}{5pt plus 2pt minus 2pt}

\begin{algorithm}[H]
\caption{$LowValuation(\T)$}\label{alg: low valuation}
    \DontPrintSemicolon
    \SetKwFunction{topYs}{XLeaves}
    
    \SetKwFunction{post}{$\texttt{BV}_{\texttt{post}}$}
    \SetKwFunction{pre}{$\texttt{BV}_{\texttt{pre}}$}
    \SetKwFunction{exist}{Projn}
    \SetKwInOut{function}{Operation}
    \SetKwInOut{Parameter}{Notation}
    \SetKwInOut{Explanation}{Explanation}

    \Explanation{This algorithm computes pre and post-valuations of nodes in the leaves and in the $Y$ partition, simultaneously in the bottom-up manner checks if the realizability set is tautology (fully realizable) or an \emph{obvious} negation (not realizable). If neither applies, it passes the new tree to Algorithm~\ref{alg: high valuation}.}
    \KwIn{$\T = (T, r, \gamma, \pi)$: an $(X,Y)$-graded project-join tree with internal nodes partitioned into $\I_X$, $\I_Y$.}
    \KwOut{Returns \emph{full} or \emph{nullary realizability} if can determine by the end of this algorithm. Otherwise, a project-join tree $\T_X$ of $(\exists_{y \in Y}y)\phi$ is passed to Algorithm~\ref{alg: high valuation} for further determining between \emph{partial} and \emph{nullary realizability}.}
    
    $B_{pureX} \gets $conjunction of clauses without $y$ variables\;
    \lIf {\upshape $B_{pureX} == \llbracket 0 \rrbracket$} {
        \Return{nullary realizable}
    }
    $\T_X \gets \T$, $b \gets 1$
    \;
    \lIf {\upshape ${\I_Y}$ is empty}{
        \Return{fully realizable}
    } 
    
    \For {$n \in \I_Y \cup \Lv{T}$}{
    {
        
        $\texttt{GenericValuation}(\T, n)$ }\tcp{in bottom-up order from leaves to root }
        \lIf {$\pre(\T_X, n)==\llbracket0\rrbracket$}{
            \Return{nullary realizable}
        }
    }
    
    \For {$n \in \topYs(\T)$}{

        \lIf {$\post(\T_X, n)==\llbracket0\rrbracket$}{
            \Return{nullary realizable}
        }
        \ElseIf{$\post(\T_X, n)\neq\llbracket1\rrbracket$}{
            $b \gets 0$\;
        }

        
        $\mathcal{V}(T_X) \gets \mathcal{V}(T_X)\setminus D(n)$\tcp{remove the descendants of $n$ from $\T_X$}
        
        
        $\Lv{T_X} \gets \Lv{T_X}\cup\{n\}$\tcp{add $n$ to leaves of $\T_X$}
    }
    \lIf{$b == 1$ and $B_{pureX} == \llbracket1\rrbracket$}{
        \Return{fully realizable}
    }
    \Return{\emph{\texttt{HighValuation}($\T_X$)}}
\end{algorithm}

\begin{theorem}\label{thm: alg:low correctness full}
Given a graded project-join tree $\T$ for a CNF formula $\phi$, Algorithm~\ref{alg: low valuation} returns full realizability if and only if the formula $(\forall X)(\exists Y)\phi(X,Y)$ is true. 
And the algorithm returns nullary realizability, if and only if the formula $(\exists X)(\exists Y)\phi(X,Y)$ is false.
\end{theorem}
\noindent When the formula is not fully realizable, we proceed to pass the tree returned by Algorithm~\ref{alg: low valuation} to Algorithm~\ref{alg: high valuation}, which then distinguishes partial realizability from nullary realizability. It does this by simply using Algorithm~\ref{alg: generic} to compute the post-valuation of the root of $\T_X$. 


In the case of the example in Figure~\ref{figure: intermediate trees}, for this formula, we do not need to check Algorithm~\ref{alg: high valuation} because by Algorithm~\ref{alg: low valuation} full realizability is returned.

\setlength{\floatsep}{5pt plus 2pt minus 2pt}
\setlength{\textfloatsep}{5pt plus 2pt minus 2pt}
\setlength{\intextsep}{5pt plus 2pt minus 2pt}

\begin{algorithm}[H]
\caption{$HighValuation(\T_X)$}\label{alg: high valuation}
    \DontPrintSemicolon
    
    \DontPrintSemicolon
    \SetKwFunction{BVpost}{$\texttt{BV}_{\texttt{post}}$}
    \SetKwFunction{exist}{Projn}
    \SetKwInOut{function}{Operation}

    \KwIn{$\T_X = (T_X, r_X, \gamma_X, \pi_X)$ is the generated tree from Algorithm~\ref{alg: low valuation}, which is the project-join tree generated by projecting out all $Y$ variables from the $(X,Y)$-graded tree for CNF $\phi$.}
    \KwOut{whether or not $\phi$ is partially realizable}

    \For {$n \in \I_X$ \tcp{for all nodes in $I_X$ partition}}{
        $\texttt{GenericValuation}(\T, n)$ \tcp{in bottom-up order leaves to the root }
    }
    \lIf {$\BVpost(\T_X, r_X)==\llbracket0\rrbracket$}{
        \Return{nullary realizable}
    } \lElse{
        \Return{partially realizable}
    }
\end{algorithm}

\noindent Since this corresponds to taking the conjunction of the leaves (which are themselves the post-valuation of the first layer of $Y$ nodes) and existentially quantifying the $X$ variables, the result is $1$ if $\varphi$ is partially realizable, and $0$ if not. Built on top of Theorem~\ref{thm: preval(n)=R(X)}, we formulate the correctness of the algorithm:

\begin{theorem}\label{thm: high valuation correctness}
Given a CNF formula $\phi$ and its graded project-join tree $\T$, if we generate $\T_X$ by Algorithm \ref{alg: low valuation}, then Algorithm \ref{alg: high valuation} returns the correct status of realizability of $\phi$.
\end{theorem}

\section{Synthesis of Witness Functions}\label{sec: synthesis(new)}

\noindent We now move to the third phase of our boolean-synthesis approach, where we construct boolean expressions for the output variables, which are the witness functions. 
More precisely, when the realizability set $R_\varphi(X)$, as defined in Section~\ref{sec: realizability(new)}, is nonempty, we proceed to witness construction. We now formally define the concept of witnesses, in the context where the boolean synthesis problem is given as a CNF specification and reduced ordered BDDs are used for boolean-function representations. 


The following lemma describes the \emph{self-substitution method} for witness construction.
\begin{lemma}{\rm \cite{CAV16}}\label{lemma: witness one y}
Let $X$ be a set of input variables and $y$ a single output variable in a Boolean CNF formula $\phi(X, y)$ with $R_\phi(X) \neq \emptyset$. Then  $g(X) = \phi(X, y)[y \mapsto 1]$ is a witness for $y$ in $\phi(X, y)$.
\end{lemma}

\noindent We now show how to extend this method to multiple output variables, building towards an approach using graded project-join trees.

\subsection{Monolithic Approach}\label{subsec: monolithic and clustering}

\noindent The synthesis procedure here builds on the condition that partial realizability is known, provided by the algorithms in Section.~\ref{sec: realizability(new)}. While solvers constructed by related works, as discussed in Section.~\ref{sec: introduction(new)}, apply only to fully realizable formulas, we show here that synthesis can also be performed to obtain witness functions in the case of partial realizability.

As a stepping stone towards graded-tree-based synthesis, we first explain how witness functions are constructed in the monolithic case. We review the basic framework in~\cite{CAV16}, where the realizability set $R_{\phi}(X)$ is obtained through iterative quantification on $y_n$ to $y_1$, while witnesses are obtained via iterative substitution on $y_1$ to $y_n$. 

Denote the BDD encoding the original CNF formula $\phi(X,Y)$ to be $B_\phi(X,Y)$. 
Then, a series of intermediate BDDs can be defined on the way of obtaining the realizability sets $R_\phi(X)$:
\setlength{\textfloatsep}{0.1cm}
\setlength{\floatsep}{0.1cm}
\allowdisplaybreaks
\begin{align*}
    &B_n(X, y_1, \ldots, y_n) \equiv B_\phi, \\
    &\ldots\\
    &B_{i}(X, y_1, \ldots, y_i)
        \equiv (\exists y_{i+1}) B_{i+1},\\
    &\ldots\\
    &B_{0}(X) 
        \equiv (\exists y_{1}) B_{1}
\end{align*}
%
Finally, the realizability set $R_\varphi(X)$ is $B_0(X)$.

Note that existential quantification proceeds here \emph{inside-out}, since larger-indexed output variables are quantified before smaller-indexed output variables.
Witness construction, on the other hand, proceeds \emph{outside-in}: witnesses are constructed in the reverse direction starting from the smallest-indexed output variable $y_1$. The witnesses for variables $y_1, \ldots, y_{i-1}$ are substituted into $B_{i}$, from which we then construct the next witness $g_i$. Using the witness given by Lemma~\ref{lemma: witness one y}, we have:

$g_1$ is computed from:
$B'_1(X,y_1)=B_1(X,y_1),$  via $g_{1}=B'_1[y_1 \mapsto 1];$

$\ldots$

$g_{i}$ is computed from: $B'_i(X,y_i)=B_i(X,y_1,\ldots,y_i)[y_{1} \mapsto g_{1}]$
$\ldots[y_{i-1}\mapsto g_{i-1}],$ via $g_{i}=B'_i(X,y_i)[y_i \mapsto 1].$


The following lemma is based on \cite{CAV16}.
\begin{lemma}\label{lemma: monolithic correct}
If $R_0(X)\not=\emptyset$, then
the $g_i$'s above are witness functions for $Y$ in $\phi(X,Y)$.
\end{lemma}

\noindent As the witnesses above are computed using the self-substitution method from Lemma~\ref{lemma: witness one y}, each formula can have potentially many different witnesses. The correctness of the procedure does not depend on which witness is used. RSynth \cite{CAV16} applies the \texttt{SolveEqn} function of the CUDD package to compute a block of witnesses at once by essentially the procedure above.  The procedure produces several witnesses for each variable, from which we chose one witnesses (by setting a parameter to 1).

The above monolithic synthesis framework was generalized in~\cite{FMCAD17} to the case where the formula $\phi$ is given as a conjunction of \emph{factors}. A factor can be a formula, for example, a single clause or a conjunction of clauses (called a ``cluster'' in that work). In that case, the principle of early quantification mentioned in Section~\ref{subsec: quantification on variables} can be applied. See Appendix B for details. As shown in \cite{FMCAD17}, the tool, \emph{Factored RSynth},  generally outperforms RSynth.



In the following section, we describe a factored approached to witness construction using graded project-join trees.

\subsection{Synthesis Using Graded Project-Join Trees}\label{subsec: graded pj trees appraoch}

\noindent As we saw above, in the monolithic setting we compute the realizability set by quantifying the $Y$ variables inside out, and then computing the witness function for these variables outside in. In the graded project-tree framework, we saw that the realizability set is computing by quantifying the $Y$ variables bottom-up. 

Our framework works for both partial and fully realizable cases. 
We first compute realizability sets going \emph{bottom up} the graded project-join tree for $\phi$ as in Section 3.
We now show that the witness functions for the $Y$ variables can then be constructed by iterated substitution \emph{top-down}. 
In other words, we first compute the witnesses for the variables in the labels of internal nodes at higher levels, and then propagate those down toward the leaves. We compute witnesses from the pre-valuation $\texttt{BV}_{\texttt{pre}}$ of a node in the tree, computed as in Algorithm~\ref{alg: generic}. 
Note that, in contrast to the bottom-up realizability and top-down synthesis described here, in projected model counting, where graded project-join trees were introduced \cite{Procount}, the trees are processed fully in a bottom-up fashion.

The following lemma is crucial to our approach.
\begin{lemma}\label{lemma: independence}
Let $m$ and $n$ be two different internal nodes of a graded project-join tree $\T$ such that $n$ is not a descendant of $m$. Then no variable in $\pi(m)$ appears in $\texttt{BV}_{\texttt{pre}}(\T, n)$.
\end{lemma}

\noindent We can derive from Lemma~\ref{lemma: independence} the essential relation among witness functions for different variables: since the witness for a variable $y_i \in \pi(n)$ is computed from $\texttt{BV}_{\texttt{pre}}(\T, n)$, this witness can only depend on the witnesses of output variables $y_j \in \pi(m)$ such that $n$ is a descendant of $m$.

Based on these insights, we present in Algorithm~\ref{alg: synth} our dynamic-programming synthesis algorithm for producing the witnesses $g_y(X)$ for each output variable $y$, represented by a BDD $W_y(X)$. The algorithm starts at the top-most layer of $Y$ nodes, those whose parents are $X$ nodes, represented by the set called \texttt{XLeaves} (line~\ref{line:xleaves}). For each node $n$ in this set (line~\ref{line:synthNodes}), we compute the set of witnesses $\{W_y \mid y \in \pi(n)\}$ from $\texttt{BV}_{\texttt{pre}}(\T, n)$ (line~\ref{line:SolveEqn}) using the monolithic procedure from Section~\ref{subsec: monolithic and clustering} (represented by the CUDD function $\texttt{SolveEqn}(\pi(n), \texttt{BV}_{\texttt{pre}}(\T, n))$, mentioned in that section). Note that, since the tree is graded, these nodes do not descend from any $Y$ nodes. Therefore, by Lemma~\ref{lemma: independence}, these witnesses depend only on the $X$ variables.

As we compute the witnesses for a node, we add all of its (non-leaf) children to the set of nodes to visit (line~\ref{line:newsynthNodes}), representing the next layer of the tree. After all nodes in the current set have been processed, we repeat the process with the new layer (line~\ref{line:synthNodes}). 

\setlength{\floatsep}{5pt plus 2pt minus 2pt}
\setlength{\textfloatsep}{5pt plus 2pt minus 2pt}
\setlength{\intextsep}{5pt plus 2pt minus 2pt}

\begin{algorithm}[H]
\caption{${DPSynth}(\T)$}\label{alg: synth}
     \DontPrintSemicolon
     \SetKwFunction{topYs}{XLeaves}
     \SetKwFunction{preBV}{pre-valuation}
     \SetKwFunction{syn}{constructWitness}
     \SetKwFunction{laterYs}{UpperYs}
     \SetKwInOut{Parameter}{Notation}

     \Parameter{$C(n)$: set of the children nodes of $n$}
     \Parameter{$D(n)$: set of the descendant nodes of $n$}
     \Parameter{$W_y$: BDD representing the witness of variable $y$}

    \Parameter{$\llbracket \alpha \rrbracket$: the BDD representation of boolean expression $\alpha$}
    
    \Parameter{$\topYs(\T)$ as in Algorithm~\ref{alg: low valuation}}
    
    \KwIn{$\T = (T, r, \gamma, \pi)$: the original $(X,Y)$-graded project-join tree of CNF $\phi$}
    \KwIn{ BDDs $\texttt{BV}_{\texttt{pre}}(m)$ for all $m \in \I_Y$}
    
    \KwOut{a series of BDDs $W_y, \forall y \in Y$, encoding the witness functions for output variables}
    
    $\texttt{synthNodes} \gets \topYs(\T)$\; \label{line:xleaves}
    \While{$\texttt{synthNodes} \neq \{\}$}{ \label{line:emptyset}
        \tcp{from top-down order from the root to leaves}
        \For{all $n \in \texttt{synthNodes}$}{ \label{line:synthNodes}
            \For{$n' \in C(n)$}{
                \If{$n'$ is not a leaf}{
                  add $n'$ to \texttt{synthNodes}\; \label{line:newsynthNodes}
                }
            }
            \tcp{monolithic algorithm presented in Section~\ref{subsec: monolithic and clustering}} 
            $\{W_y \mid y \in \pi(n)\} \gets \texttt{SolveEqn}(\pi(n), \texttt{BV}_{\texttt{pre}}(\T, n))$\; \label{line:SolveEqn}

            \For{$n'' \in D(n)$}{\label{line:subst_begin}
                \For{all $y \in \pi(n)$}{
                    \tcp{substitute new synthesized $y$ by their witnesses $W_y$ in the BDD representing the pre-valuation of descendants $n''$}
                    $\texttt{BV}_{\texttt{pre}}(\T, n'') \gets \texttt{BV}_{\texttt{pre}}(\T, n'')[y \mapsto W_{y}]$\;
                }
            }\label{line:subst_end}
        }
        
    }
    \Return{$W_y, \forall y \in Y$} \label{line:return}

\end{algorithm}

\noindent This continues until the set of children is empty (line~\ref{line:emptyset}). Note that, since for each node $n$ we apply the monolithic synthesis procedure only to the variables in $\pi(n)$, the witnesses can be dependent on the $Y$ variables of its ancestors. Therefore, we finish the algorithm by iterating over the new synthesized witnesses $W_y$ and substituting each into the pre-valuation of descendant nodes that are computed later (lines~\ref{line:subst_begin}-\ref{line:subst_end}). Note that this overapproximates the set of dependencies on $y \in \pi(n)$ of the witnesses for $y' \in \pi(n'')$ for $n'' \in D(n)$, but since $W_{y'}[y \mapsto W_y] \equiv W_{y'}$ when $y$ does not appear in $W_{y'}$, the result is still correct. At the end of this procedure, all $W_y$ will be dependent only in the input variables $X$.

Continuing with the running example for the problem in Figure~\ref{figure: intermediate trees}, once full realizability is detected, we apply Algorithm~\ref{alg: synth} and construct the witnesses in top-down manner. First, we get the witness $g_6=1$ for $y_6$ by $\texttt{BV}_{\texttt{pre}}(\T, 9)= \llbracket(\neg x_1 \lor x_2 \lor y_6)\rrbracket$. Then we construct the witness $g_5=(x_1 \land x_2) \lor \neg x_3$ for $y_5$ by substituting $y_5=1$ in $\texttt{BV}_{\texttt{pre}}(\T, 7)=\llbracket(x_1 \lor \neg x_2 \lor x_3 \lor y_5) \land (\neg x_3 \lor x_2 \lor \neg y_5)  \land (x_1 \lor \neg y_5 \lor \neg x_3)\rrbracket$. After that, we go to node $6$ and get $g_4 = x_1 \lor \neg x_3$ from $\texttt{BV}_{\texttt{pre}}(\T, 6)=\llbracket (x_1 \lor y_4 \lor \neg y_5) 
\land (\neg x_3 \lor x_1 \lor \neg y_4) \rrbracket$. By the algorithm, we would need to substitute $g_5$ into $\texttt{BV}_{\texttt{pre}}(\T, 6)$ before computing $g_4$, but since for this specific CNF $g_4$ is not actually dependent on $g_5$, this does not change the witness. The witnesses are correct by $\phi[y_4 \mapsto g_4][y_5 \mapsto g_5][y_6 \mapsto g_6]=1$.

The following theorem proves the correctness of witnesses constructed. First, it is easy to see by an inductive argument that a witness is synthesized for all output variables: all $Y$ nodes that do not descend from other $Y$ nodes are included in the set processed in the first iteration, and if a node is processed in one iteration, all of its children are included in the set for the next iteration. Since the tree is graded, we have processed all of the $Y$ nodes, and since every output variable $y$ is in the label of some node, a witness $W_y$ is computed for every $y$. Then, the following theorem states the correctness of the witnesses constructed:



 
\begin{theorem}\label{thm: alg synth}
Algorithm~\ref{alg: synth} returns a set of BDDs encoding the witness functions for output variables that satisfy the given CNF $\varphi$, assuming that $\texttt{SolveEqn}(\pi(n)$, $\texttt{BV}_{\texttt{pre}}(\T, n))$ returns correct witnesses for the variables in $\pi(n)$ in $\texttt{BV}_{\texttt{pre}}(\T, n)$.
\end{theorem}


\section{Experimental Evaluation}\label{sec:experiments}

\subsection{Realizability-Checking Phase}\label{subsec: real experi}

{\bf Methodology}:
To examine our dynamic-programming  graded-project-join-tree-based approach for boolean realizability, we developed a software tool, \emph{DPSynth}, which implements the theoretical framework described above. 
We choose to compare \emph{DPSynth} to \emph{Factored RSynth}, which as explained in the introduction is the closest existing tool, also being based on decision diagrams and outputting the realizability set along with the witnesses.
Prior work~\cite{FMCAD17} already demonstrated that RSynth practically never outperforms Factored RSynth, while Factored RSynth typically outperforms RSynth, so we compare in this paper to Factored RSynth. (See Section~\ref{sec: manthan} for an additional comparison to a non-decision-diagram tool.)
We measured the time and space performance for determining realizability on a set of mature benchmarks, described below. 
The experiments aim at answering the following research questions: 
\begin{itemize}
\item Does our DP-based solution improve execution time for realizability checking? Does the overhead of \emph{planning} time investment in DPSynth get paid off? 
\item How does the relative weight of planning overhead vary between small and large input instances?
\item How does DPSynth improve the scalability of realizability checking?
\end{itemize}
(Tree width is one of the critical parameters impacting the performance of the graded project-jointree-based approach in running time. We discuss this issue further in separate a subsection.)

We selected 318 benchmarks that are neither too easy (taking less than 1ms) not too hard (such that the whole benchmark family is not solvable by either solver) from the data set of forall-exists $\Pi^P_2$ CNFs from the QBFEVAL~\cite{qbfeval} datasets from 2016 to 2019, without any more selection criteria, as the editions of QBFEVAL in 2020 and 2022 do not include additional 2QBF instances or tracks other than those included by 2019. Families of benchmarks that our experiments run on include \emph{reduction-finding query, mutexP, qshifter, ranking functions, sorting networks, tree and fix-point detection} families, and also two additional scalable families, consisting of parametric integer factorization and subtraction benchmarks from \cite{akshay2021boolean,supratik2}.

Among the full benchmark suite, at least $33\%$ of the $285$ instances where realizability can be checked by either DPSynth or factored RSynth are partially realizable. 
In the benchmarks for which both tools are able to synthesize a complete group of witnesses, $28\%$ are partially realizabile.
(No benchmark in the suite is identified as nullary realizable).
We conclude that partial realizability is a significant issue in Boolean synthesis.
We now present the experimental results analyzed according to the research questions above

For each benchmark instance, we run the FlowCutter-based planner until we obtain the first tree decomposition, and we declare timeout if no tree decomposition is generated within ten minutes, in which case, the instance is marked unsolved. Otherwise, we take the first tree generated, and proceed to the execution phase of DPSynth, which includes BDD compilation, realizability checking, and synthesis of witnesses. The maximal time limit for each instance is set to be two hours for the execution phase. We measure planning time, execution time, and end-to-end time. 

Our implementation is based on the CUDD Library~\cite{CUDD_cudd} with BDD operations, and the FlowCutter tree-decomposition tool~\cite{flowcutter}.
In our implementation, BDD variables are in MCS order \cite{FMCAD17}. This ordering is based on the primal graph of the input clause set, and was also used by Factored RSynth. 

We ran the experiments on Rice University NOTS cluster, which assigns the jobs simultaneously to a mix of HPE SL230s, HPE XL170r, and Dell PowerEdge C6420 nodes, each of which has 16-40 cores with 32-192 GB RAM that runs at 2.1-2.60 GHz. Each solver-benchmark combination ran on a single core.

{\bf Experimental Results}:
 Our goal in this paper was to compare the performance of the fast, CSP-based, formula-partitioning techniques of Factored RSynth \cite{FMCAD17}, to the heavier-duty formula-partitioning techniques based on tree decomposition described above. For a fair comparison, the charts do not include data points for those instances that timed out for one of the solvers.  (In total, DPSynth was able to solve 126 instances end-to-end, and RSynth 111.) Thus, our experiments compare \emph{DPSynth} to Factored RSynth.  Our conjecture is that such heavier-duty techniques pay off for larger formulas, but not necessarily for smaller formulas. 

\setlength{\floatsep}{5pt plus 2pt minus 2pt}
\setlength{\textfloatsep}{5pt plus 2pt minus 2pt}
\setlength{\intextsep}{5pt plus 2pt minus 2pt}
\begin{figure}[H]
\centering

\begin{subfigure}{0.48\textwidth}\centering
\includegraphics[width=\textwidth]{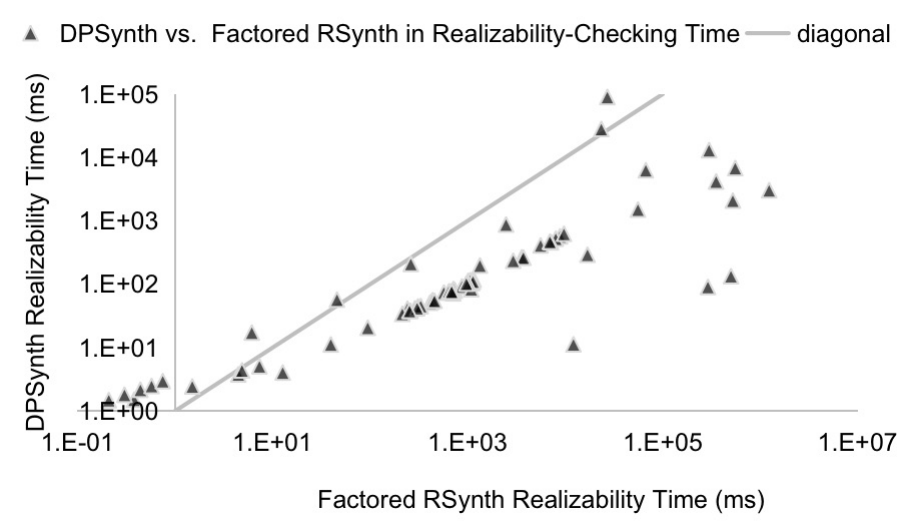}
\caption{Time Comparison} 
\label{figure: new realiz time compar}
\end{subfigure}
\hfill
\begin{subfigure}{0.48\textwidth}\centering
\centering
\includegraphics[width=\textwidth]{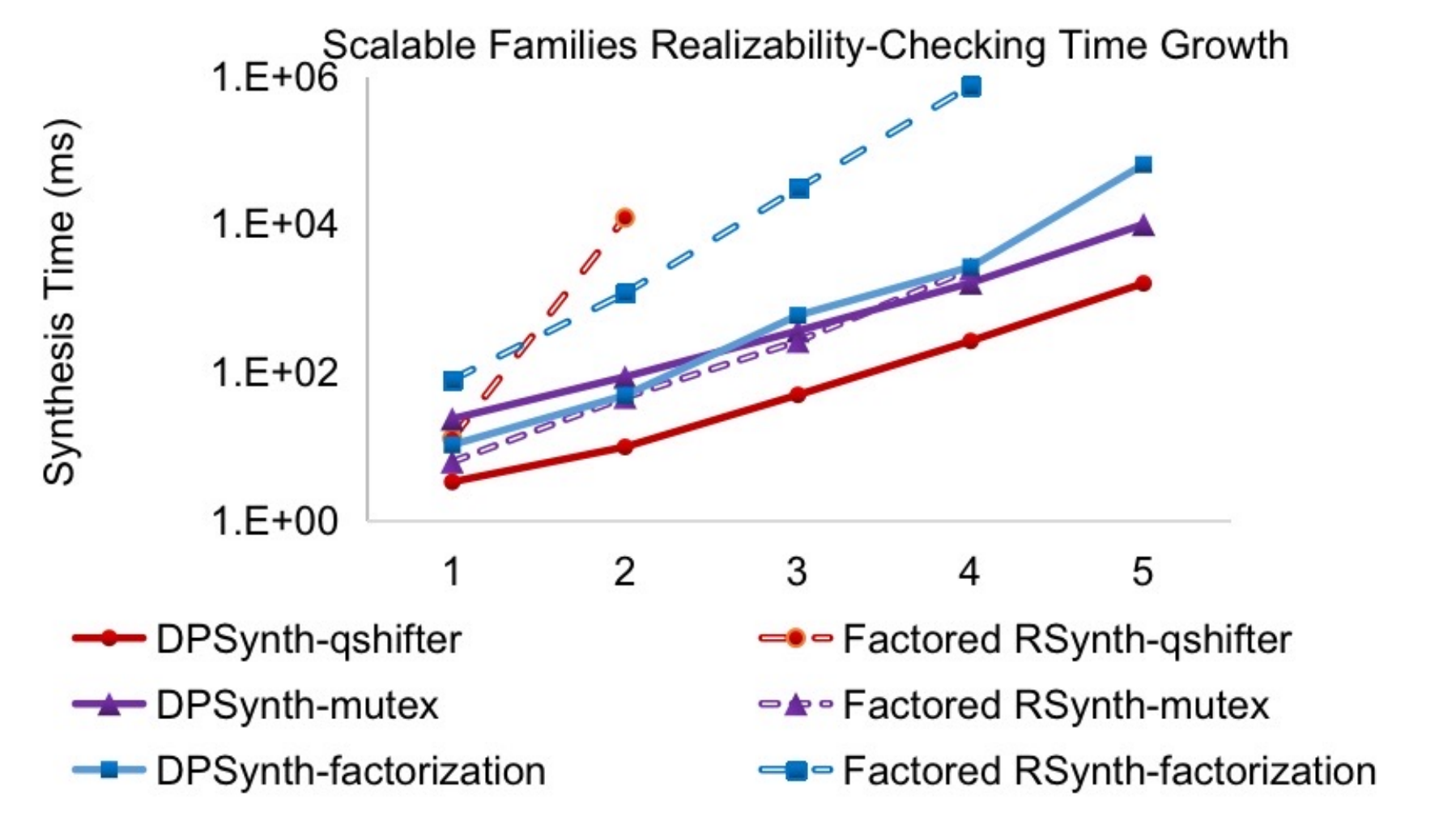}
\caption{Scalable Families Time Growth} \label{figure: scalable realiz}
\end{subfigure}

\caption{} 
\end{figure}

\noindent Figure~\ref{figure: new realiz time compar} shows overall running-time comparison between DPSynth and Factored RSynth for realizability checking. A clear pattern that emerges is there is a difference in relative performance between very small problems (solvable within $1$ millisecond) and larger problems. DPSynth underperforms Factored RSynth  on small instances, but outperforms on larger instances and the difference increases exponentially as input size grows. We conclude that planning overhead dominates on small input instances, but that effect fades off as instances get larger and the planning pays off.
In memory usage comparison\protect\footnote{The charts for space consumption for both phases are in the Appendix C.}
-- using peak node count as a measurement -- DPSynth uses less memory than Factored RSynth. Here the graded project-join-tree approach is advantageous, and planning incurs no overhead.



To evaluate scalability, we take the scalable benchmark families mentioned previously, and compare the logarithmic-scale slope in their running time as sizes of benchmarks increase. As indicated in Figure.~\ref{figure: scalable realiz}, DPSynth scales exponentially better than Factored RSynth, as the planning overhead fades in significance as instances grow. We see a steeper slope on the exponential scale in DPSynth trends over factored RSynth. Some data points for larger benchmarks in the families (horizontal coordinates $3,4,5$ on chart) are missing because factored RSynth is not able to finish solving these instances in realizability-checking phase within the time limit.

\subsection{Synthesis}\label{subsec: synthesis experiments}

The experiments on synthesis of witnesses answer the following research questions: 
\begin{itemize}
\item How does DPSynth compare in execution time to factored RSynth? \item Does the influence of \emph{planning} investment reduces as problem gets large?
\item What do we see in growth of tree widths and synthesis execution time?
\end{itemize}

\begin{figure}[H]
\centering
\begin{subfigure}{0.48\textwidth}\centering
\includegraphics[width=\textwidth,trim=4 4 4 4,clip]{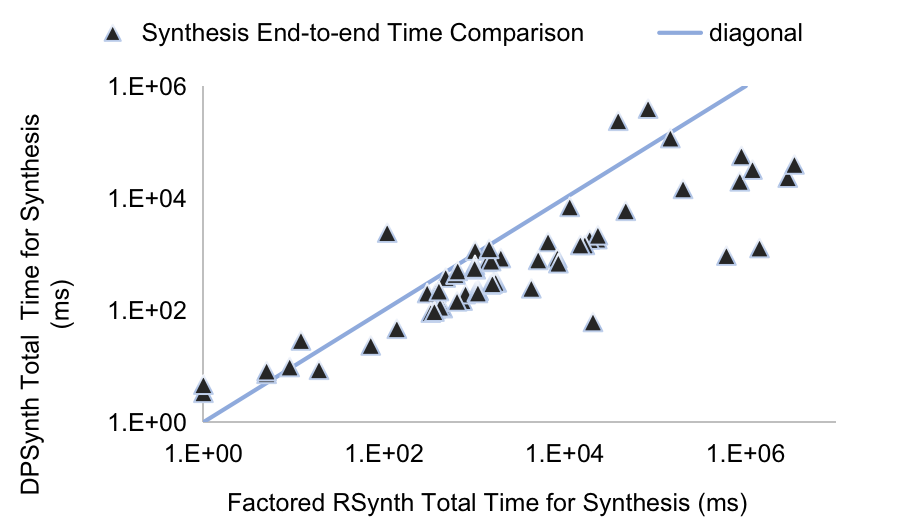}
\caption{Synthesis Time: DPSynth vs. Factored RSynth} 
\label{figure: new synth compar}
\end{subfigure}
\hfill
\begin{subfigure}{0.52\textwidth}\centering
\includegraphics[width=\textwidth,trim=4 4 4 4,clip]{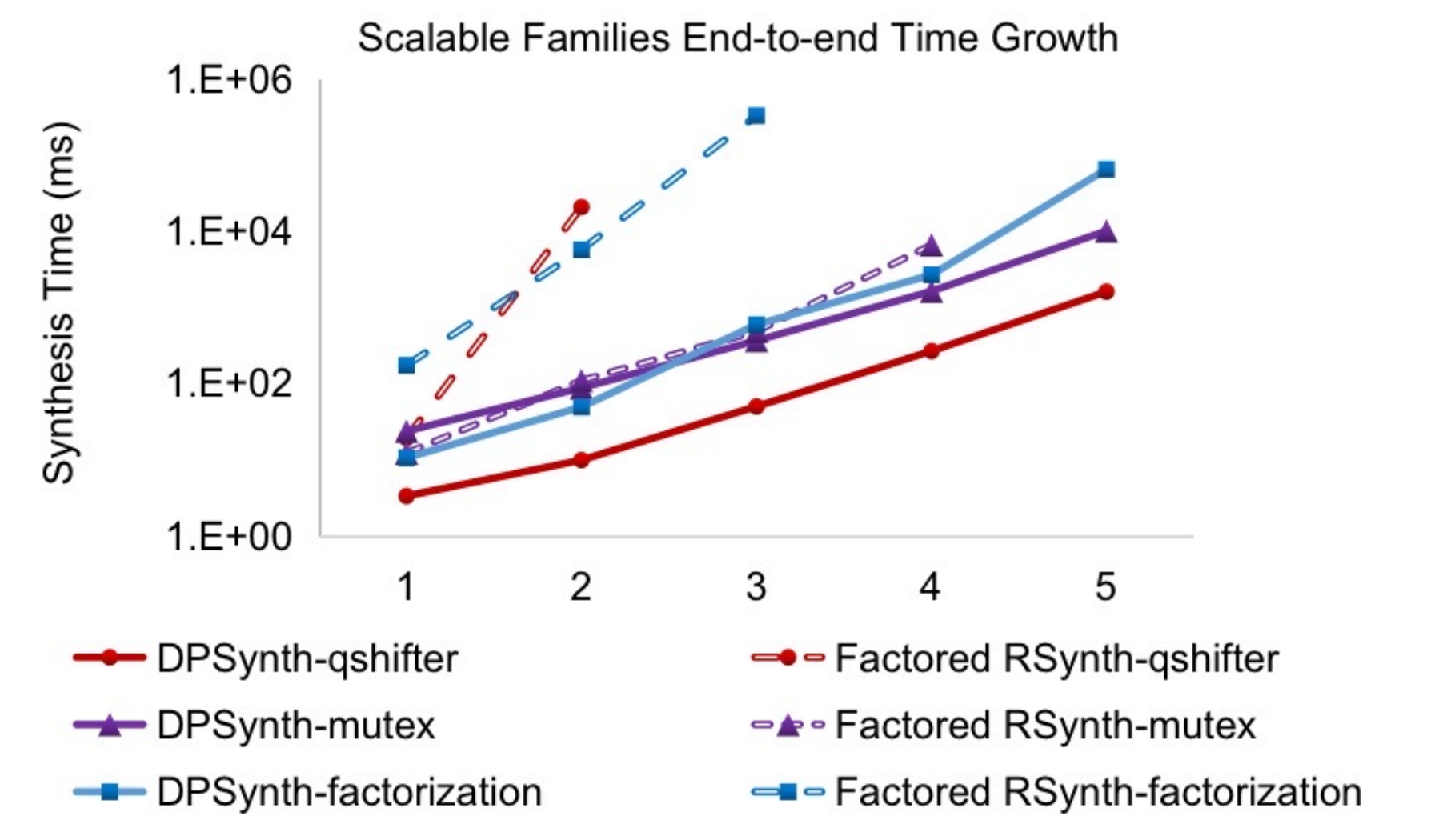}
\caption{Scalable Families Comparison} \label{figure: scalable synth}
\end{subfigure}
\caption{}
\end{figure}

Using the same set of benchmarks and setting under the methodology as in Section~\ref{subsec: real experi}, we applied our synthesis procedure to both fully realizable and partially realizable benchmarks. This broadens the scope of Boolean synthesis beyond that of fully realizable benchmarks, which is the scope of earlier work, as discussed above. 
Regarding end-to-end synthesis (planning, realizability checking, and synthesis, combined) DPSynth outperforms Factored RSynth in running time as illustrated in Figure~\ref{figure: new synth compar}. As with realizability checking, planning overhead dominates for small instances, as discussed in Section~\ref{sec: realizability(new)}, but DPSynth solves large benchmarks faster and shows significant advantage as problem size increases, once the planning overhead fades in significance. In memory usage there is consistent relative performance of DPSynth vs. Factored RSynth.



We again selected three scalable benchmark families, scaled based on a numerical parameter. As shown in Figure~\ref{figure: scalable synth}, DPSynth scales exponentially as benchmark size increases. Similarly to the case in realizability checking, the missing data points on larger benchmarks in the scalable families presented by the dashed lines are caused by the timeouts by factored RSynth.

While DPSynth shows performance advantage over Factored RSynth with respect to our benchmark suite, one cannot conclude that DPSynth always dominates Factored RSynth. This is because DPSynth involves computationally nontrivial planning phase, and it is not possible to say definitively that the planning overhead always pays off. 

\noindent We can conclude, however, that DPSynth is an important addition to the portfolio of algorithms for boolean synthesis.

\subsection{Tree Widths and Realizability}

Graded project-join trees enable the computation of the realizability set in a way that minimizes the set of support of intermediate ADDs (Algebraic Decision Diagrams), saving time and memory. But computing these trees is a heavy computational task. In this section, we study the impact of tree width on realizability checking.

\begin{figure}[H]\centering

\begin{subfigure}{0.48\textwidth}
\centering
\includegraphics[width=\textwidth]{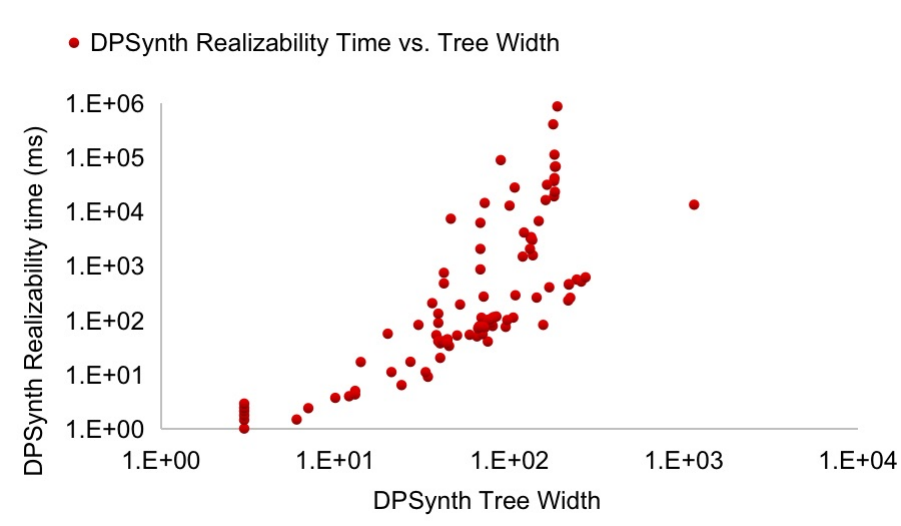}
\caption{DPSynth Realizability Time as Widths Increases} \label{figure: new DPSynth realiz time vs width}
\end{subfigure}
\hfill
\begin{subfigure}{0.48\textwidth}
\centering
\includegraphics[width=\textwidth]{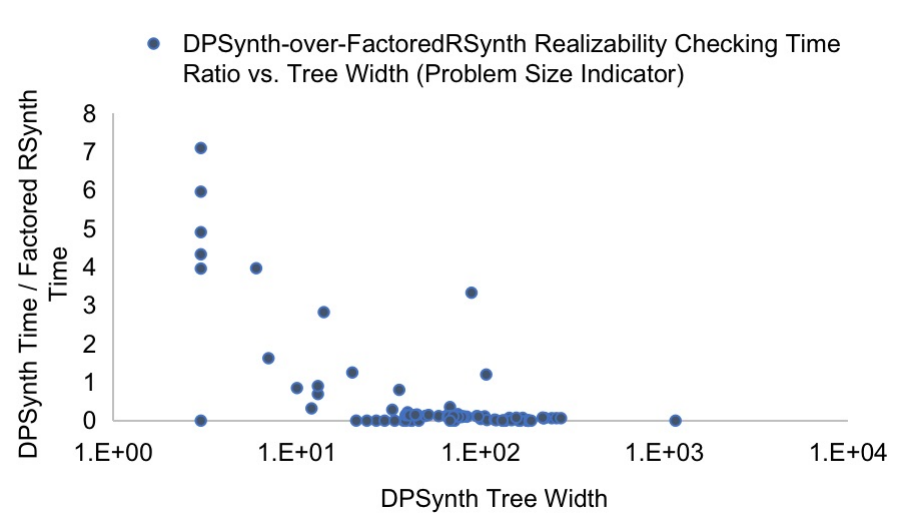}
\caption{DPSynth / Factored RSynth Realizability Time Ratio as Widths Increases} \label{figure: new realiz time ratio vs width}
\end{subfigure}
\caption{}
\end{figure}
\noindent Figure~\ref{figure: new DPSynth realiz time vs width} presents the range of realizability time run by DPSynth along increasing tree widths.  As we see,
increases in \emph{tree widths} implies an increase in running time for realizability-checking. Figure~\ref{figure: new realiz time ratio vs width}, depicts the ratio of realizability time, computed by that of DPSynth over Factored RSynth, with respect to tree width, for problem instances that can be solved by both solvers. As treewidth increases, the over-performance of  DPSynth over Factored RSynth increases, as planning overhead decreases in significance for higher-treewidth instances.






Synthesis execution time has similar relation with tree widths.

\subsection{Comparison with Non-BDD-Based Synthesis} \label{sec: manthan}

\noindent To complement out evaluation, we include additional experiments to verify whether DPSynth is competitive with non-BDD-based synthesis solvers, as motivated in Section~\ref{sec: introduction(new)}. We compare with Manthan~\cite{golia2020manthan}, a leading tool that is not based on decision diagrams, and find that DPSynth performs favorably.

We present a general picture by the following table, which shows the overall strength of the dynamic-programming decision-diagram approach, by measuring the time and space usage of experiments on the dataset selected from QBFEVAL'16 to QBFEVAL'19. DPSynth and Manthan each is able to solve some benchmarks that the other does not solve. 



In order to compare the running time, we compare against Manthan using the DPSynth \emph{end-to-end synthesis} time, which is the sum of tree-decomposition time, compilation, realizability-checking time, and synthesis time.

As the overall picture of synthesis solving, DPSynth shows a better time performance on most instances. DPSynth and Manthan each has strength on some instances, but the number of benchmarks that only DPSynth solves is larger than those solved only by Manthan. On those that are solved by both, DPSynth takes less time in most of them. See Appendix D for more illustrative data on specific fully and partially-realizable benchmarks.

\begin{table}[H]
\centering
\begin{tabular}{|l|l|l|l|} 
 \hline
         & Number of benchmarks \\ \hline
        solved by Manthan & 79  \\ \hline
        solved by DPSynth & 102  \\ \hline
        solved by both Manthan and DPSynth & 70  \\ \hline
        solved by Manthan but not by DPSynth & 9  \\ \hline
        solved by DPSynth but not by Manthan & 32  \\ \hline
\end{tabular}
\end{table}


\section{Concluding Remarks}\label{sec:conclusions}

\noindent 
To summarize the contribution in this work, we propose a novel symbolic dynamic programming approach for realizability checking and witness construction in boolean synthesis, based on graded project-join trees. The algorithm we propose here combine a bottom-up realizability-checking phase with a top-down synthesis phase. We demonstrated experimentally that our approach, implemented in the  DPSynth tool, is powerful and more scalable than the approach based on CSP heuristics (Factored RSynth). 
Another crucial contribution of this work is the inclusion of partial realizability checking, which applies to 30\% of the total number of benchmarks. As we explain in introduction, this consideration is motivated by the need in modular circuit design and temporal synthesis to locate the scope of realizable inputs in iterative constructions.

There are many directions for future work. Variable ordering is a critical issue in decision-diagram algorithms, and should be explored further in the context of our approach. In particular, dynamic variable ordering should be investigated \cite{rudell1993dynamic}. Also, in our work here we used high-level API of the BDD package CUDD, but it is possible that performance gains can be obtained by using also low-level BDD-manipulating APIs. The question of how to provide certificates of unrealizability also needs to be explored, c.f.,~\cite{bryant2021generating,bryant2021dual}.


As mentioned earlier, quantifier elimination is a fundamental algorithmic component in temporal synthesis \cite{ZhuTLPV17}. Tabajara and Vardi explored a factored approach for temporal synthesis \cite{tabajara2019partitioning}. It would be worthwhile to explore also the graded project-tree approach for boolean synthesis in the context of factored temporal synthesis. Finally, quantifier elimination is also a fundamental operation in symbolic model checking \cite{burch1992symbolic} and partitioning techniques have been explored in that context~\cite{Burch91}. Therefore, exploring the applicability of our dynamic-programming approach in that setting is also a promising research direction.



\subsubsection{\ackname} 
This work was supported in part by NSF grants IIS-1527668, CCF-1704883, IIS-1830549, CNS-2016656, DoD MURI grant N00014-20-1-2787, an award from the Maryland Procurement Office, the Big-Data Private-Cloud Research Cyberinfrastructure MRI-award funded by NSF under grant CNS-1338099 and by Rice University's Center for Research Computing (CRC).

\newpage
\bibliographystyle{splncs04}
\bibliography{main}
\newpage

\appendix
\section{Proofs} \label{appendix: proofs}

\subsection{Proof of Theorem.~\ref{thm: alg.generic correctness}}
\noindent
\begin{proof}
We first prove the correctness of post-valuations computed.
\emph{Case 1}: If $n$ is a leaf in $\T$, $n \in \Lv{T}$, the returned $\texttt{BV}_{\texttt{post}}(\T, n)$ encodes $\gamma(n)$ in line 5, which is the clause represented by $n$. This is the same BDD as $\texttt{BV}_{\texttt{post}}(\T, n)$ in Definition~\ref{def: BDD valuation}.

\emph{Case 2}: Otherwise, we just need to show in the loop in line 9-12, BDD $\texttt{BV}_{\texttt{post}}(\T, n)$ is always equivalent to the conjunction  $ \bigwedge_{n' \in C(n)} \texttt{BV}_{\texttt{pre}}(\T, n')   $. As in Def.~\ref{def: BDD valuation}, we denote the represented boolean formula of BDD $B$ by $\langle B \rangle$. We use induction and assume that by the end of every iteration, $\langle\texttt{BV}_{\texttt{pre}}(\T, n)\rangle \equiv \bigwedge_{n'' \in C'(n)} \langle \texttt{BV}_{\texttt{post}}(\T, n'') \rangle$, where $C'(n)$ is the set of all children nodes that have been traversed by the end of the last iteration. In the base case, $\texttt{BV}_{\texttt{pre}}(\T, n) = 1 \land \texttt{BV}_{\texttt{post}}(\T, n')$, where $n'$ is the first children node traversed. Hence it is returns as defined. Specifically, if $\texttt{BV}_{\texttt{pre}}(\T, n')==\llbracket0\rrbracket$, then the iteration is the last iteration since conjunction of boolean formulas with $0$ is also $0$.

Next we prove the induction step. If in the new iteration the BDD $\texttt{BV}_{\texttt{pre}}(\T, n')$ is not $0$, then it is updated to the conjunction of $ (\bigwedge_{n'' \in C'(n)}  \texttt{BV}_{\texttt{post}}(\T, n'') ) \land \texttt{BV}_{\texttt{pre}}(\T, n')$, exactly matching the inductive hypothesis. So this invariant is preserved in this new iteration.
If in the new iteration of $n'$, the BDD $\texttt{BV}_{\texttt{pre}}(\T, n')$ encodes $0$, the conjunction in Def.~\ref{def: BDD valuation} is also $0$, so the pre-valuation line 12 is consistent with the inductive hypothesis.
Hence by the end of every iteration of the loop in lines 9-12, we have the invariant that $\texttt{BV}_{\texttt{pre}}(\T, n) \equiv \bigwedge_{n'' \in C'(n)} \texttt{BV}_{\texttt{post}}(\T, n'')$. Line 13 performs the operation of existential quantification in Def.~\ref{def: BDD valuation}. Hence we have the correctness of Alg.~\ref{alg: generic}.

Next we show the equivalence between the returned pre-valuations and its definition. For leave nodes $n$, the pre-valuation computed in line 4 corresponds to Definition.~\ref{def: BDD valuation}. In the case of internal node, $\texttt{BV}_{\texttt{pre}}(\T, n) $ is returned either by the end of the for-loop in lines 9-12, or returned whenever it is equal to $\llbracket 0 \rrbracket$. We also know that $\texttt{BV}_{\texttt{post}}(\T, n) $ is proved above to be equivalent to its definition, and the fact that a conjunction with zero is always 0, in the sense of both boolean formulas and BDDs. Therefore we are only left to show that the pre-valuation defined in Definition.~\ref{def: BDD valuation} is equivalent to $\texttt{pre-BV}(\T, n)$ at the end of Algorithm.~\ref{alg: generic}. Since $\texttt{pre-BV}(\T, n)$ is updated in line 10 only, it is equivalent to the conjunction of $\llbracket 1 \rrbracket$ and $ \texttt{BV}_{\texttt{post}}(\T, n') $ for all $n' \in C(n)$, which is exactly as defined in its Definition.~\ref{def: BDD valuation}. Hence we complete the proof.

\end{proof}

\subsection{Proof of Theorem.~\ref{thm: alg:low correctness full}}
\noindent
\begin{proof}
First, we prove the first statement of Theorem~\ref{thm: alg:low correctness full}.
Suppose $(\exists_{y \in Y} \hspace{0.1cm}y) \phi=1$, i.e., there is full realizability. We need to show that $b=1$ throughout the for-loop in lines 10-16. i.e., We need to prove that in all iterations, the algorithm never goes to line 14. That is, to show that $\texttt{BV}_{\texttt{post}}(\T_X, n)=1$ for all $n \in \texttt{XLeaves}(\T)$. By Definition~\ref{def: BDD valuation} and Theorem~\ref{thm: alg.generic correctness}, the post-valuation of $Y$ internal nodes is equal to the conjunction of post-valuations under it with its labelled $Y$ variables existentially quantified. This is then equivalent to the conjunctions of all leaves in its descendants, projecting out all variables labelled in its descendants. Since we assume full realizability, all clauses represented by the leaves are satisfiable, and thus the post-valuations of all the $Y$ internal nodes are equal to $1$, not to mention the visited internal nodes in the uppermost level of $\I_Y$.

To prove the other direction, assume that Algorithm~\ref{alg: low valuation} returns \emph{fully realizable}. Then line 17 must confirms, and it is equivalent to say $\texttt{BV}_{\texttt{post}}(\T_X, n)$ is always $1$. Then for all internal nodes in $\texttt{XLeaves}(\T)$, post-valuation is $\llbracket1\rrbracket$. Since all the leaves in the original graded project-join tree $\T$ has one and only one ancestor in $\texttt{XLeaves}(\T)$, the conjunction of the post-valuations of all $n \in \texttt{XLeaves}(\T)$ is equivalent to $(\exists_{y \in Y} \hspace{0.1cm}y) \phi $ and to $1$. Hence $\phi$ achieves full realizability.

Next, we prove that a returned \emph{not realizable} from Algorithm~\ref{alg: low valuation} is sufficient condition for null realizability of $\phi$.
Suppose Algorithm~\ref{alg: low valuation} returns \emph{not realizable}. Then in line 11, $\texttt{BV}_{\texttt{post}}(\T_X, n)=\llbracket0\rrbracket$. There is some $Y$ internal node $n$ without parents in $\I_Y$, such that its post-valuation in $\T_X$ is the $0$-BDD. Hence, by Def.~\ref{def: BDD valuation}, the conjunction of post-valuations of its children node in $\I_Y$ is $0$. Thus there does not exist $Y$ variables labelled by this node $n$, such that the conjunction of clauses in its descendants is satisfiable. Thus, there does not exist a set of boolean values for all the $Y$ variables, that satisfies the conjunction of clauses in descendants of $n$. Moreover, the conjunction of all clauses in $\phi$ is not satisfiable. Thus, $\phi$ does not achieve partial realizability in its domain and null realizability is proved. 

Hence we complete the proof for Theorem~\ref{thm: alg:low correctness full}.

\end{proof}

\subsection{Proof of Theorem.~\ref{thm: preval(n)=R(X)}}
\noindent
This theorem follows from the definition of realizability set and Definition.~\ref{def: BDD valuation}:
\begin{proof}
    Let $\T$ be a graded project-join tree of CNF $\phi(X,Y)$. Let $X_\texttt{leaves}(\T)$ be the set of highest $Y$ nodes. Denote descendants of node $n$ by $D(n)$.
    
    From Definition.~\ref{def: BDD valuation},
    
    $$BV_\texttt{pre}(\T, n) \equiv (\exists_{z \in V } z)\bigwedge_{l \in D(n) \cap \Lv{\T}} \gamma(l)$$ where $V = \{z \mid \exists m \in D(n) s.t. z \in \pi(m)\}$.
    Then the realizability set $R$ is defined to be $$R=\{(x_1 \ldots x_m) \mid (\exists y_1)\ldots(\exists y_n) s.t. \phi(x_1, \ldots, x_m, y_1, \ldots, y_n)=1\}=(\exists_{\forall y \in Y} y)\bigwedge_{l \in \Lv{\T}} \gamma(l).$$

    We also know that each $Y$ variable is labelled under one and only one highest level $Y$ node, by definition of a graded project-join tree. Hence $$Y = \bigcup_{n \in X_\texttt{leaves}(\T)} \left( \bigcup_{m \in D(n)} \pi(m) \right). $$

    Therefore, the conjunction of pre-valuations for all $n \in X_\texttt{leaves}(\T)$, i.e., $\bigwedge_{n \in {X_\texttt{leaves}}(\T)} \texttt{BV}_{\texttt{pre}}(\T, n)$, we can show:

    \begin{align*}
        &\bigwedge_{n \in {X_\texttt{leaves}}(\T)} \texttt{BV}_{\texttt{pre}}(\T, n)
        \\\equiv &\bigwedge_{n \in {X_\texttt{leaves}}(\T)} (\exists_{z \in V } z)\bigwedge_{l \in D(n) \cap \Lv{\T}} \gamma(l) \texttt{ where } V = \{z \mid \exists m \in D(n) s.t. z \in \pi(m)\}
        \\\equiv &(\exists_{z \in V' } z) \bigwedge_{n \in {X_\texttt{leaves}}(\T)} \bigwedge_{l \in D(n) \cap \Lv{\T}} \gamma(l) \\&\texttt{ where } V' = \{z \mid \exists n \in {X_\texttt{leaves}}(\T), \exists m \in D(n) s.t. z \in \pi(m)\}\\ &\tag{ since a variable only occurs in clauses represented a leaves under its ancestors}\\ \tag{by definition of graded project-join tree}
        \\\equiv &(\exists_{z \in V' } z) \bigwedge_{n \in {X_\texttt{leaves}}(\T)} \bigwedge_{l \in D(n) \cap \Lv{\T}} \gamma(l) \\&\texttt{ where } V' = \{z \mid \exists m \in \I_Y s.t. z \in \pi(m)\}\\ &\tag{by definition of $\I_Y$}
        \\\equiv &(\exists_{z \in Y } z) \bigwedge_{n \in {X_\texttt{leaves}}(\T)} \bigwedge_{l \in D(n) \cap \Lv{\T}} \gamma(l)  \\ &\tag{by definition of $\T$}
        \\\equiv &(\exists Y) \bigwedge_{n \in {X_\texttt{leaves}}(\T)} \bigwedge_{l \in D(n) \cap \Lv{\T}} \gamma(l) 
        \\\equiv &(\exists Y_\phi)\bigwedge_{l \in \Lv{\T}} \gamma(l). \\ \tag{by definition of $\T$ and notation in the statement to be proved}
    \end{align*}
    Theorem~\ref{thm: preval(n)=R(X)} is proved.
\end{proof}

\subsection{Proof of Theorem.~\ref{thm: high valuation correctness}}
\noindent
\begin{proof}
Suppose Algorithm~\ref{alg: high valuation} returns \emph{not realizable}. Then the post-valuation of the root is $0$-BDD. By Definition~\ref{def: BDD valuation}, it is either a leave representing an unsatisfiable clause, or the $X$ variables do not satisfy the conjunction of all clauses in the tree. In the trivial former case it is obviously not realizable. 

In the latter case, since $\T_X$ is the output of Algorithm \ref{alg: low valuation} where all $Y$ internal nodes are eliminated, the post-valuation of these new leaves remain the same in the updated tree $\T_X$. 
Taking existential projections on an output variable only to the clauses where it has occurrences in, is logically equivalent to quantifying it on the entire formula. The same reasoning applied to their symbolic representations. Hence the conjunction of the clauses in $\T_X$ with all $X$ variables quantified is logically equivalent to $\varphi$ with all $X$ and $Y$ existentially quantified. Hence, there do not exist $X$ and $Y$ variables that satisfy $\varphi$. i.e., if Algorithm~\ref{alg: high valuation} returns \emph{not realizable}, there is null realizability.

On the other hand, assume Algorithm~\ref{alg: high valuation} returns \emph{partially realizable}. Then in line 1 $\texttt{BV}_{\texttt{post}}(\T_X, r_x)$ is not $0$-BDD. That is, the conjunction of leaves has satisfying set of variables labelled. By analysis above, it is equivalent to satisfiability of the CNF formula represented by the original tree $\T$ in Algorithm~\ref{alg: low valuation}. That is, there is partial realizability. 
Since null and partial realizability is complementary to each other, we do not need to prove the reverse directions in both cases. Hence the correctness of Theorem~\ref{thm: high valuation correctness} is proved.
\end{proof}

\noindent

\noindent


\subsection{Proof of Lemma.~\ref{lemma: independence}}
\noindent
\begin{proof}
From the definition of of project-join tree, if $y \in \pi(m)$ then every clause where $y$ appears must be in a leaf node that descends from $m$. Since $n$ is not a descendant of $m$, then either no clauses containing $y$ have been joined on the way to form $\texttt{BV}_{\texttt{pre}}(\T, n)$, or (if $m$ is a descendant of $n$ instead) $y$ has already been existentially quantified.
\end{proof}

\subsection{Proof of Theorem.~\ref{thm: alg synth}}
\noindent
\begin{proof}
We use $C(n)$ to denote the set of children of node $n$, $D(n)$ the set of descendants of node $n$ (not including $n$ itself), and $L(n) \subseteq D(n)$ the set of descendants of $n$ that are leaves.
We also use the notation $Y_n = Y \cap \pi(n)$ for the set of variables synthesized in node $n$ and $Y'_n = \bigcup_{n' \in D(n)} Y_n$ for the set of variables synthesized in nodes that are descendants of $n$. 
Finally, we denote by $W^n_y = W_y[y' \mapsto W_{y'}]_{y' \in Y_n \cup Y'_n}$ the witness for variable $y$ after all witnesses synthesized in $n$ or its descendants have been substituted into it. Note that if $y \in Y_m$ where $m$ is not a descendant of $n$ (including when $m = n$) then $W^n_y = W_y$ (by Lemma~\ref{lemma: independence}).

Assume that, for an arbitrary node $n$, $\texttt{SolveEqn}(\pi(n), \texttt{BV}_{\texttt{pre}}(\T, n))$ returns a correct witness $W_y$ for every $y \in \pi(n)$. In other words,
that $\texttt{BV}_{\texttt{pre}}(\T, n)[y \mapsto W_y]_{y \in Y_n} \equiv (\exists Y_n) \texttt{BV}_{\texttt{pre}}(\T, n)$. Recall that both $W_y$ and $\texttt{BV}_{\texttt{pre}}(\T, n)[y \mapsto W_y]_{y \in Y_n}$ can still be dependent not only on the $X$ variables, but also on the $Y$ variables of nodes that $n$ descends from (but no other nodes, as per Lemma~\ref{lemma: independence}).

We show by induction that the following holds for every node $n$:
\begin{align*}
\label{eq:inductive}
&\left(\bigwedge_{c \in L(n)} \gamma(c)\right)[y \mapsto W^n_y]_{y \in Y_n}\\
 &[y \mapsto W^n_y]_{y \in Y'_n} \equiv (\exists Y_n) (\exists Y'_n) \bigwedge_{c \in L(n)} \gamma(c).
\end{align*} 

This expression states that $\{W^n_y \mid y \in Y_n \cup Y'_n\}$ are witnesses for $\bigwedge_{c \in L(n)} \gamma(c)$ (note that these witnesses might still depend on $Y$ variables from ancestors of $n$). If $n = r$ is the root of the tree, then $\bigwedge_{c \in L(r)} \gamma(c) \equiv \varphi$ and $Y_r \cup Y'_r = Y$, and so $\varphi[y \mapsto W^r_y]_{y \in Y} \equiv (\exists Y) \varphi$. In other words, $\{W^r_y \mid y \in Y\}$ are witnesses for the $Y$ variables. Note that the $W^r_y$ are equivalent to the final witnesses returned in line~\ref{line:return}. Therefore, the witnesses returned by Algorithm~\ref{alg: synth} are correct. All that is left is to prove~(\ref{eq:inductive}).

\paragraph{Base case.} Assume $n$ is a leaf. In that case, $L(n)$, $Y_n$ and $Y'_n$ are all empty, and $\bigwedge_{c \in L(n)} \gamma(c) \equiv 1$. Therefore, (\ref{eq:inductive}) holds vacuously.

\paragraph{Inductive step.} Assume $n$ is an internal node, and (\ref{eq:inductive}) holds for all of $n$'s children $C(n) = \{n_1, \ldots, n_k\}$. We will show that it holds for $n$ as well. Note that we use the fact that, by early quantification, $\texttt{BV}_{\texttt{pre}}(\T, n) \equiv (\exists Y'_n) \bigwedge_{c \in L(n)} \gamma(c)$.
\allowdisplaybreaks
\begin{align*}
  & (\exists Y_n) (\exists Y'_n) \bigwedge_{c \in L(n)} \gamma(c) \\
        \equiv& (\exists Y_n) \texttt{BV}_{\texttt{pre}}(\T, n)\\
          &  \tag{since $\texttt{BV}_{\texttt{pre}}(\T, n) \equiv (\exists Y'_n) \bigwedge_{c \in L(n)} \gamma(c)$} \\
        \equiv& \texttt{BV}_{\texttt{pre}}(\T, n)[y \mapsto W^n_y]_{y \in Y_n}\\
          &  \tag{by correctness of \texttt{SolveEqn}} \\ \tag{and by $W^n_y = W_y$} \\
        \equiv& \left((\exists Y'_n) \bigwedge_{c \in L(n)} \gamma(c)\right)[y \mapsto W^n_y]_{y \in Y_n}\\
          &  \tag{since $\texttt{BV}_{\texttt{pre}}(\T, n) \equiv (\exists Y'_n) \bigwedge_{c \in L(n)} \gamma(c)$} \\
        \equiv& \left(\bigwedge^k_{i = 1} (\exists Y_{n_i}) (\exists Y'_{n_i}) \bigwedge_{c \in L(n_i)} \gamma(c) \right) \\
        & [y \mapsto W^n_y]_{y \in Y_n}\\
          &  \tag{by early quantification} \\ & \tag{and by the fact that the $L(n_i)$ are disjoint} \\
        \equiv& \left(\bigwedge^k_{i = 1} \left(\bigwedge_{c \in L(n_i)} \gamma(c)\right)[y' \mapsto W^{n_i}_{y'}]_{y' \in Y_{n_i} \cup Y'_{n_i}} \right)\\
        &[y \mapsto W^n_y]_{y \in Y_n}
            \tag{by induction hypothesis} \\
        \equiv& \left(\bigwedge^k_{i = 1} \bigwedge_{c \in L(n_i)} \gamma(c) \right)[y' \mapsto W^{n_1}_{y'}]_{y' \in Y_{n_1} \cup Y'_{n_1}}\ldots \\
        &[y' \mapsto W^{n_k}_{y'}]_{y' \in Y_{n_k} \cup Y'_{n_k}}[y \mapsto W^n_y]_{y \in Y_n} 
            \tag{since the $L(n_i)$ are disjoint} \\
        \equiv& \left(\bigwedge^k_{i = 1} \bigwedge_{c \in L(n_i)} \gamma(c) \right)[y \mapsto W^n_y]_{y \in Y_n} \\
        &[y' \mapsto W^{n}_{y'}]_{y' \in Y_{n_1} \cup Y'_{n_1}}\ldots [y' \mapsto W^{n}_{y'}]_{y' \in Y_{n_k} \cup Y'_{n_k}} 
            \tag{*} \label{step:star} \\
        \equiv& \left(\bigwedge^k_{i = 1} \bigwedge_{c \in L(n_i)} \gamma(c) \right)[y \mapsto W^n_y]_{y \in Y_n}[y' \mapsto W^{n}_{y'}]_{y' \in Y'_n} 
            \tag{since $Y_{n_1} \cup Y'_{n_1} \cup \ldots \cup Y_{n_k} \cup Y'_{n_k} = Y'_n$} \\
        \equiv& \left(\bigwedge_{c \in L(n)} \gamma(c) \right)[y \mapsto W^n_y]_{y \in Y_n}[y' \mapsto W^{n}_{y'}]_{y' \in Y'_n}.
            \tag{since $L(n_1) \cup \ldots \cup L(n_k) = L(n)$}
\end{align*}

The step marked with~(\ref{step:star}) above merits a more detailed explanation. In the step before, the substitution $[y \mapsto W^n_y]_{y \in Y_n}$ is performed at the end, and therefore this substitution also affects the witnesses $W^{n_i}_{y'}$ from the child nodes (which might depend on the variables $y \in Y_n$). When we move this substitution to the beginning, to ensure equivalence we must also substitute $W^n_y$ into the substitutions $[y' \mapsto W^{n_i}_{y'}]_{y' \in Y_{n_i} \cup Y'_{n_i}}$, which updates $W^{n_i}_{y'}$ to $W^{n_i}_{y'}[y \mapsto W^n_y]_{y \in Y_n} \equiv W^{n}_{y'}$.
\end{proof}


\section{Synthesis in Factored RSynth} \label{subsec: factored synthesis}
\noindent

Let $\phi = F_1 \land \ldots \land F_m$ (each $F_j$ is a factor, for example a clause or a cluster). We associate with the factor $F_j$ the set $Y_j$ of the output variables that occur in $F_j$ but do not occur in a factor $F_{i}$ with $i<j$. Equivalently speaking, the set $Y_j$ consists of the variables from $F_j$ that only occur in factors $F_j, \ldots, F_m$. Moreover, since each $Y_j$ includes all variables that satisfy this criterion, for all $1 \leq j \leq m$, if a variable $y \in Y_j$, then it is guaranteed that $y\not\in Y_k$ for $k\neq j$.

The intuition behind this definition is that an output variable $y\in Y_j$ does not depend on the formulas $F_1,\ldots,F_{j-1}$. Thus, these clauses can be ignored when constructing a witness for $y$. i.e., the values of clauses in these other factors do not affect the function constructed for this variable. 
We use the notation $\mathcal{F}_j$ to denote the BDD representing the factor $F_j$, and $\exists Y_j$ to denote the existential quantification of all variables $y \in Y_j$ as a block. We can then compute realizability by using early quantification:
\begin{align*}
    &B_{m-1} \equiv (\exists Y_m) \mathcal{F}_m \\
    &B_{m-2} \equiv (\exists Y_{m-1}) (\mathcal{F}_{m-1} \land B_{m-1}) \\
    &\ldots \\
    &B_i \equiv (\exists Y_{i+1}) (\mathcal{F}_{i+1} \land B_{i+1}) \\
    &\ldots \\
    &B_1 \equiv (\exists Y_2) (\mathcal{F}_2 \land B_2) \\
    &B_0 \equiv (\exists Y_1) (\mathcal{F}_1 \land B_1) 
\end{align*}
Note that, from the correctness of early quantification, we have that $B_i \equiv (\exists Y_{i+1})\ldots(\exists Y_m)(\mathcal{F}_{i+1} \land \ldots \mathcal{F}_m)$. Like before, $B_0$ is equivalent to the realizability set $R_\phi$. Note that this framework already has some kind of tree structure. Each $\mathcal{F}_j$ corresponds to a leaf, and each $B_j$ corresponds to an internal node, with children $\mathcal{F}_{j-1}$ and $B_{j-1}$, and labeled by $Y_{j-1}$. $B_0$ is the root of the tree. Similarly to how realizability was computed inside-out in the monolithic case, in this case realizability is computed \emph{bottom-up}, from the leaves to the root. Witness construction, which was computed outside-in in the monolithic case, is in this case computed \emph{top-down}, from the root to the leaves (witnesses in a single node can be computed as in the monolithic procedure):
\begin{align*}
    &\textit{For all $y_i \in Y_1$, 
    $g_i$ is computed from} \\
    &\hspace{0.1cm}(\mathcal{F}_1 \land B_1), \\
    &\textit{For all $y_i \in Y_2$, 
    $g_i$ is computed from} \\
    &\hspace{0.1cm}(\mathcal{F}_2 \land B_2)[y_j \mapsto g_j]_{y_j \in Y_1}, \\
    &\ldots \\
    &\textit{For all $y_i \in Y_k$, $g_i$ is computed from} \\
    &\hspace{0.1cm}(\mathcal{F}_k \land B_k)[y_j \mapsto g_j]_{y_j \in Y_1 \cup \ldots \cup Y_{k-1}}, \\
    &\ldots \\
    &\textit{For all $y_i \in Y_{m-1}$, $g_i$ is computed from} \\
    &\hspace{0.1cm}(\mathcal{F}_{m-1} \land B_{m-1})[y_j \mapsto g_j]_{y_j \in Y_1 \cup \ldots \cup Y_{m-2}}, \\
    &\textit{For all $y_i \in Y_m$, $g_i$ is computed from} \\
    &\hspace{0.1cm}(\mathcal{F}_m \land B_m)[y_j \mapsto g_j]_{y_j \in Y_1 \cup \ldots \cup Y_{m-1}}.
\end{align*}

Note that, similarly to how in the monolithic case each witness depends on the witnesses of the previous variables, in the factored case the witnesses of a child depend on the witnesses of its ancestors. Parent nodes propagate witnesses to their children, who then substitute them into $(\mathcal{F}_k \land B_k)$ before computing witnesses for their own variables. These witnesses are then propagated to their own children. Note that $(\mathcal{F}_k \land B_k)$ corresponds to the pre-valuation from Definition~\ref{def: BDD valuation}. 


\section{Space Performance Charts}\label{sec: appendix space charts}

\noindent
\begin{figure}[H]\centering
\begin{subfigure}{0.46\textwidth}\centering
\includegraphics[width=\textwidth]{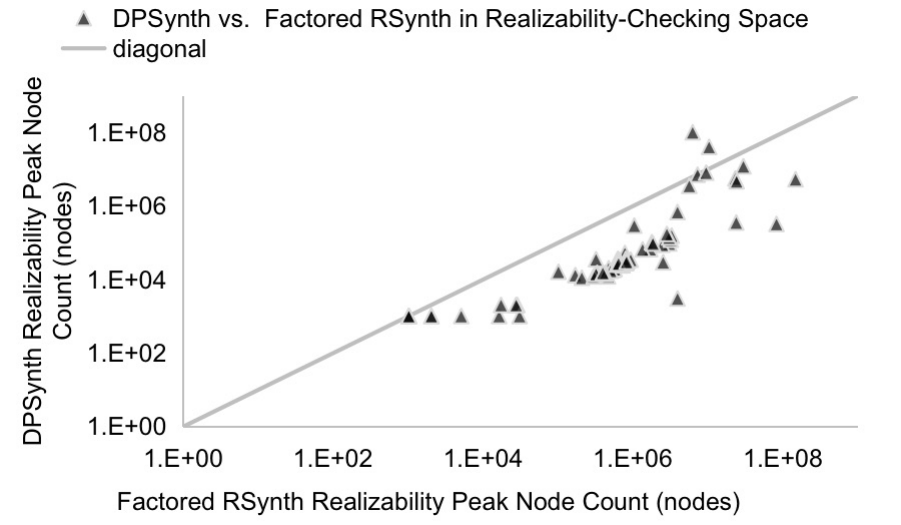}
\caption{Realizability-Checking Phase Space Usage Comparison} 
\label{figure: new realiz space compar}
\end{subfigure}
\hfill
\begin{subfigure}{0.5\textwidth}\centering
\includegraphics[width=\textwidth]{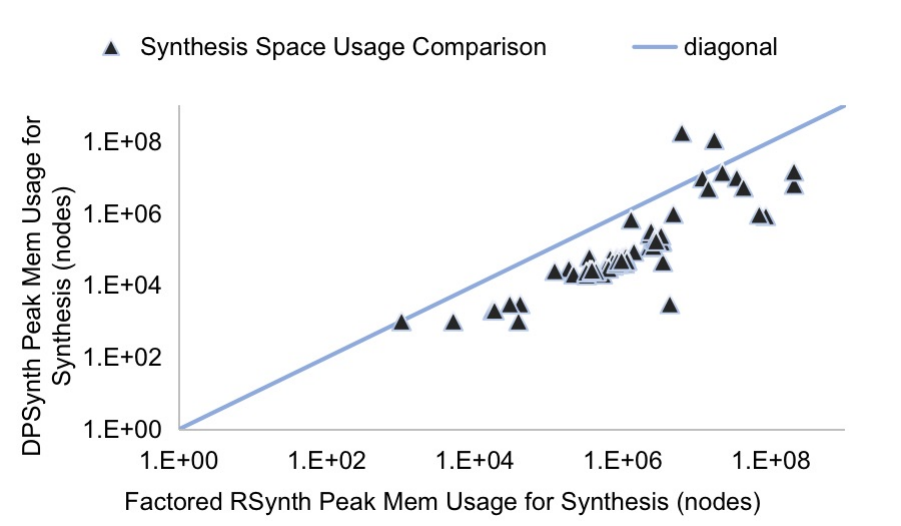}
\caption{Synthesis Space Usage Comparison} 
\label{figure: new synth space}
\end{subfigure}
\caption{}
\end{figure}

\section{DPSynth vs. Manthan}\label{appendix: manthan}
\setlength{\arrayrulewidth}{0.2mm}
\setlength{\tabcolsep}{5pt}
\renewcommand{\arraystretch}{1.2}
    
We measure the time and space usage of experiments on the dataset selected from QBFEVAL'16 to QBFEVAL'19. DPSynth and Manthan~\cite{golia2020manthan} each solves for some benchmarks that the other does not solve. 

As the overall picture of synthesis solving, DPSynth and Manthan each has strength on some instances, but the number of benchmarks that only DPSynth solves is larger than those solved only by Manthan. On the intersection of solved sets, DPSynth takes less time usage.

Overall, DPSynth shows a better time performance. In order to compare the running time, we compare against Manthan using the DPSynth \emph{end-to-end synthesis} time, which is the sum of tree-decomposition time, compilation, realizability-checking time, and synthesis time. For \emph{partially realizable} instances, since Manthan returns UNSAT for partially realizable benchmarks, we compare with Manthan by measuring the \emph{realizability time} of DPSynth, which is the sum of tree-decomposition, compilation and realizability-checking time. This is illustrated in the table presented in Section~\ref{sec:experiments} of formal text.

The results show a better performance in DPSynth running time, in both fully and partially realizable cases. This shows the strength of the dynamic-programming decision-diagram approach.

\begin{table}[H]
    \centering
    \begin{tabular}{|l|l|l|l|}
    \hline
Benchmark & Manthan time (s) & DPSynth time (s) & Manthan / DPSynth time \\ \hline
mutex-2-s       & 26.62365198      & 0.017328861      & 1536.376337                                                      \\ \hline
mutex-4-s       & 70.10412574      & 0.057425382      & 1220.786407                                                      \\ \hline
qshifter\_3     & 0.02080822       & 0.004227581      & 4.922015666                                                      \\ \hline
qshifter\_4     & 0.035684586      & 0.011800355      & 3.024026444                                                      \\ \hline
qshifter\_5     & 0.118482351      & 0.056386868      & 2.101240156                                                      \\ \hline
qshifter\_6     & 0.548010349      & 0.290776052      & 1.884647465                                                      \\ \hline
qshifter\_7     & 3.304552555      & 1.692991177      & 1.951901817                                                      \\ \hline
stmt1\_145\_146 & 1.481198549      & 0.007940356      & 186.5405719                                                      \\ \hline
stmt1\_20\_21   & 1.161505938      & 0.005571153      & 208.4857367                                                      \\ \hline
stmt1\_60\_61   & 138.0251365      & 1.923284128      & 71.7653385                                                       \\ \hline
stmt1\_629\_630 & 60.86966228      & 0.820687097      & 74.16914742                      
\\ \hline

stmt1\_787\_788 & 92.2114253       & 1.340908081      & 68.76789439                                                      \\ \hline
stmt1\_811\_812 & 5.000648737      & 0.042724069      & 117.045236                                                       \\ \hline

stmt1\_919\_920 & 16.3737123       & 0.216405419      & 75.6622102                                                       \\ \hline
stmt11\_643\_645  & 13.83426762 & 0.148434139 & 93.20138689 \\ \hline

stmt124\_966\_965 & 17.15093589 & 0.226107464 & 75.8530284  \\ \hline

stmt16\_0\_1      & 0.078123331 & 0.021049437 & 3.711421406 \\ \hline

stmt16\_285\_286  & 12.3903234  & 0.199766912 & 62.02390214 \\ \hline
stmt16\_47\_48    & 127.7006507 & 1.851167077 & 68.98386012 \\ \hline
stmt16\_818\_819  & 8.039676428 & 0.099669646 & 80.66323851 \\ \hline
stmt16\_950\_951  & 21.96217847 & 0.321454774 & 68.32120797 \\ \hline
stmt17\_99\_98    & 147.940861  & 2.241282086 & 66.00724733 \\ \hline
stmt18\_258\_260  & 9.160577059 & 0.098143359 & 93.33873583 \\ \hline
stmt2\_480\_551   & 16.77611017 & 0.226805481 & 73.9669522  \\ \hline
stmt2\_649\_647   & 11.21149755 & 0.139258954 & 80.50827055 \\ \hline
stmt2\_649\_723   & 11.58830237 & 0.14303965  & 81.01461639 \\ \hline
stmt2\_649\_776   & 14.4783535  & 0.196231457 & 73.7820211  \\ \hline
stmt21\_326\_327  & 22.5036993  & 0.357303104 & 62.98209853 \\ \hline
stmt24\_148\_149  & 1.197834015 & 0.006430307 & 186.2794443 \\ \hline
stmt24\_292\_293  & 9.379323483 & 0.109969875 & 85.28993493 \\ \hline
stmt24\_7\_8      & 0.017041206 & 0.001758292 & 9.691909171 \\ \hline
stmt24\_765\_766  & 18.09966111 & 0.220442702 & 82.10596653 \\ \hline
stmt27\_296\_297  & 10.50176525 & 0.115043276 & 91.28534597 \\ \hline
stmt27\_946\_955  & 20.3260808  & 0.247080786 & 82.26491881 \\ \hline
stmt3\_639\_640   & 8.168511868 & 0.079512754 & 102.7320959 \\ \hline

\end{tabular}
\end{table}
\begin{table}[H]
    \centering
    \begin{tabular}{|l|l|l|l|}
    \hline
Benchmark & Manthan time (s) & DPSynth time (s) & Manthan / DPSynth time \\ \hline

stmt32\_570\_572  & 23.86345649 & 0.32441781  & 73.55778799 \\ \hline
stmt37\_941\_942  & 129.7759182 & 1.517532121 & 85.51774058 \\ \hline
stmt38\_943\_942  & 117.9836717 & 1.531322152 & 77.04693066 \\ \hline
stmt41\_118\_131  & 9.311798096 & 0.22846856  & 40.75745956 \\ \hline
stmt41\_566\_580  & 44.67906404 & 0.83791215  & 53.32189542 \\ \hline
stmt41\_738\_737  & 17.30380273 & 0.214580042 & 80.64031756 \\ \hline
stmt41\_738\_749  & 36.58995533 & 0.674991152 & 54.2080518  \\ \hline
stmt44\_435\_436  & 12.76951456 & 0.144744236 & 88.22123018 \\ \hline
stmt44\_554\_555         & 22.60296774          & 0.29673466  & 76.17232088 \\ \hline
stmt44\_554\_604         & 23.51070952          & 0.307722947 & 76.40219799 \\ \hline
stmt44\_916\_917         & 16.47536898          & 0.19140868  & 86.07430434 \\ \hline
stmt7\_33\_34            & 14.49155951          & 0.934846089 & 15.50154584 \\ \hline
stmt70\_191\_213         & 8.378497839          & 0.09022127  & 92.86610396 \\ \hline
stmt70\_495\_501         & 22.71755934          & 0.314747484 & 72.1770959  \\ \hline
stmt70\_854\_859         & 16.48648357          & 0.203281798 & 81.10162216 \\ \hline
stmt72\_696\_721         & 17.04460955          & 0.204169481 & 83.4826511  \\ \hline
stmt82\_224\_225         & 10.50310063          & 0.105692302 & 99.37431994 \\ \hline
stmt86\_889\_890         & 21.63464522          & 0.284072161 & 76.15897717 \\ \hline
stmt9\_350\_351          & 8.726649284          & 0.092352257 & 94.49308082 \\ \hline
subtraction32            & 3.983768463          & 285.7623599 & 0.013940844 \\ \hline
tree-exa10-10            & 0.580000877          & 0.001643786 & 352.8445171 \\ \hline
tree-exa10-15            & 0.744118929          & 0.002058999 & 361.3983926 \\ \hline
tree-exa10-20            & 0.94502902           & 0.002538212 & 372.3207598 \\ \hline
tree-exa10-25            & 1.136503696          & 0.002991756 & 379.8784715 \\ \hline
tree-exa10-30            & 1.398799896          & 0.00363048  & 385.2933761 \\ \hline
factorization8         & 26.02147222 & 0.01355423  & 1919.804534 \\ \hline
rankfunc37\_signed\_16 & 2.976966858 & 0.020192158 & 147.4318326 \\ \hline
stmt137\_903\_911      & 18.31892133 & 0.2235378   & 81.94999382 \\ \hline
stmt19\_3\_214         & 67.52295566 & 970.6645171 & 0.069563639 \\ \hline
stmt19\_66\_214        & 84.54595137 & 148.8763251 & 0.567893863 \\ \hline
stmt2\_976\_999        & 3228.63344  & 57.85746008 & 55.80323498 \\ \hline
stmt21\_5\_134         & 47.65692377 & 14.28987617 & 3.335013067 \\ \hline
stmt21\_84\_215        & 88.41444421 & 135.9611531 & 0.650291956 \\ \hline
stmt27\_16\_97         & 53.51791215 & 9.510231213 & 5.627403893 \\ \hline
stmt27\_584\_603       & 24.56984329 & 0.329176383 & 74.64035867 \\ \hline

    \end{tabular}
\end{table}

\vfill


\end{document}